\documentclass[10pt,conference]{IEEEtran}
\usepackage{cite}
\usepackage{amsmath,amssymb,amsfonts}
\usepackage{algorithmic}
\usepackage{graphicx}
\usepackage{textcomp}
\usepackage[table]{xcolor}
\usepackage[hyphens]{url}
\usepackage{hyperref}

\usepackage{booktabs}
\usepackage{tabularx}
\usepackage{makecell}
\usepackage{xspace}
\usepackage{subcaption}
\usepackage{enumitem}
\usepackage{multirow}
\usepackage{pifont}
\usepackage{cleveref}

\newcommand{\cmark}{\textcolor{green!60!black}{\ding{51}}}
\newcommand{\xmark}{\textcolor{red}{\ding{55}}}

\definecolor{jcred}{HTML}{e31a1c}
\definecolor{jcgreen}{HTML}{33a02c}
\definecolor{jcblue}{HTML}{1f78b4}
\definecolor{jcorange}{HTML}{ff7f00}
\definecolor{jcpurple}{HTML}{6a3d9a}
\definecolor{jcbrown}{HTML}{b15928}
\newcommand{\best}[1]{\textcolor{jcgreen}{\bf #1}}

\newcommand{\llamaIIIPLUSBIG}{\textsc{Llama-}\scalebox{0.9}{3.3-70B}\xspace}

\definecolor{darkblue}{RGB}{0, 70, 140}

\title{MemExplorer: Navigating the Heterogeneous Memory Design Space for Agentic Inference NPUs}

\author{
\IEEEauthorblockN{
Haoran Wu$^{*}$\quad
Zeyu Cao$^{*}$\quad
Yao Lai$^{*}$\quad
Binglei Lou$^{\dagger}$\quad
Jiayi Nie$^{*}$\quad
Can Xiao$^{\dagger}$\quad
Timi Adeniran$^{*}$\quad
Kevin Lau$^{\dagger}$\\
Przemyslaw Forys$^{\dagger}$\quad
Kauser Johar$^{\ddagger}$\quad
Catriona Wright$^{\ddagger}$\quad
Junyi Liu$^{\S}$\quad
Kai Shi$^{\S}$\quad
Nicholas D. Lane$^{*}$\\
Rika Antonova$^{*}$\quad
Jianyi Cheng$^{\P}$\quad
Timothy Jones$^{*}$\quad
Aaron Zhao$^{\dagger}$\quad
Robert Mullins$^{*}$}
\IEEEauthorblockA{
$^{*}$University of Cambridge\quad
$^{\dagger}$Imperial College London\quad
$^{\ddagger}$Chipletti\quad
$^{\S}$Microsoft\quad
$^{\P}$University of Edinburgh}
}

\begin{document}
\maketitle
\thispagestyle{plain}
\pagestyle{plain}

\begin{abstract}

Emerging agentic large language model (LLM) workloads are driving the rapidly growing demand on both memory capacity and bandwidth, with different phases of inference (e.g., prefill and decode) imposing distinct requirements. Industry is responding by composing heterogeneous accelerators into single interconnected systems, as exemplified by NVIDIA's Vera Rubin platform, where each device brings its own memory architecture.
This heterogeneity is compounded by a widening landscape of memory technologies available: high-density on-chip SRAM, HBM, LPDDR, GDDR, and emerging options such as high-bandwidth flash (HBF), each offering their own capacity, bandwidth, and power trade-offs.
Identifying efficient memory architectures for next-generation inference accelerators remains challenging due to the vast and rapidly evolving design space across workload characteristics, NPU design dimensions, and memory system designs

To address this challenge, we present \textbf{MemExplorer}, a new memory system synthesizer for heterogeneous NPU systems.
MemExplorer provides a unified abstraction for modeling diverse memory technologies across the different hierarchies (e.g., on-chip and off-chip) and automatically determines an efficient heterogeneous memory system together with NPU design choices (e.g., matrix engine size) to balance throughput and power across prefill and decode devices in a multi-device system.

Experimental results show that, under the same power budget for agentic workloads, MemExplorer achieves up to $2.3\times$ higher energy efficiency than the baseline NPU and $3.23\times$ over H100 in the prefill-only setting. Under equivalent performance targets in the decode setting, it delivers up to $1.93\times$ and $2.72\times$ higher power efficiency over the baseline NPU and H100, respectively.

\end{abstract}

\section{Introduction}

\begin{figure}[htbp]
    \centering
    \captionsetup[subfigure]{font=normal, skip=2pt}
    \captionsetup{font=normal, skip=3pt}
    \begin{subfigure}{\linewidth}
        \centering
        \includegraphics[width=\linewidth]{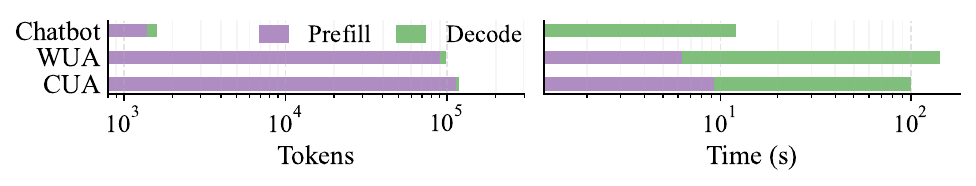}
        \caption{Agentic workloads exhibit significantly larger token processing requirements, particularly during the prefill stage.\footnotemark[1]}
        \label{fig:motivation_agentic}
    \end{subfigure}
    \vspace{0.5em}
    \begin{subfigure}{\linewidth}
        \centering
        \includegraphics[width=\linewidth]{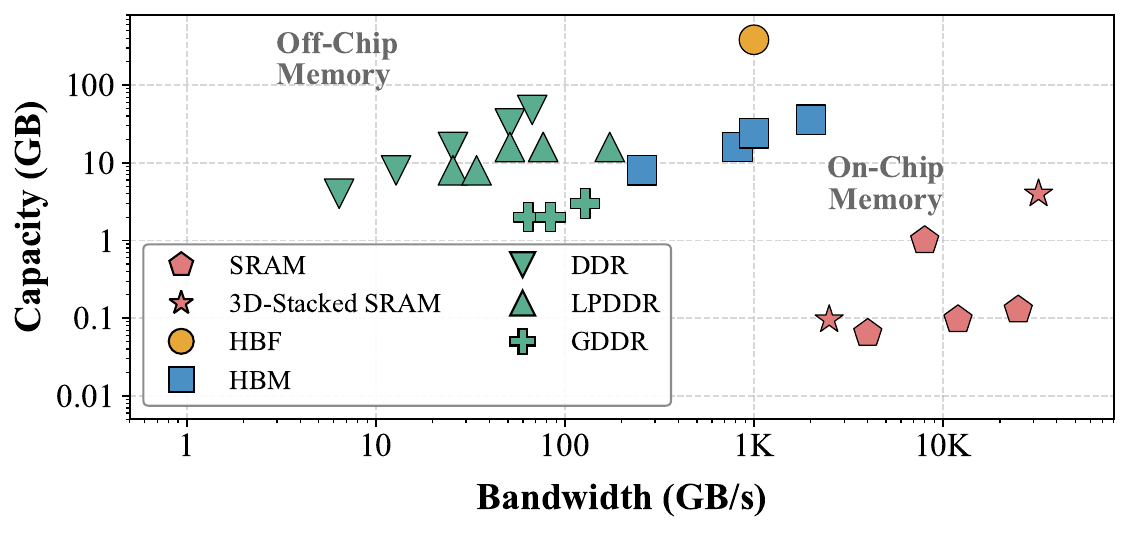}
        \caption{The growing diversity of memory technologies in capacity, bandwidth, and access latency makes the memory design space increasingly complex.}
        \label{fig:mem_design_space}
    \end{subfigure}
    \vspace{0.5em}
    \begin{subfigure}{\linewidth}
        \centering
        \includegraphics[width=\linewidth]{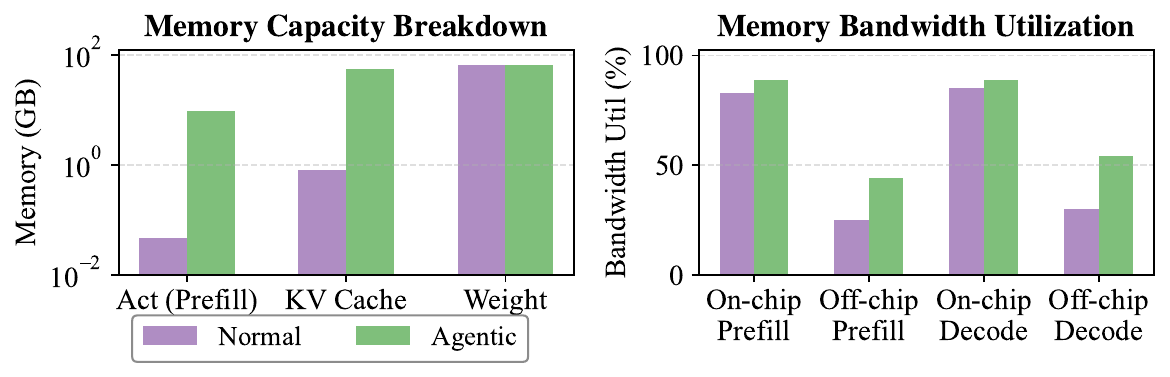}
        \caption{Agentic workloads sharply increase capacity demand for activations and KV cache, while bandwidth rises only moderately and differs between prefill and decode. The two scale independently, motivating memory technologies that provision them separately rather than a single HBM tier.\footnotemark[2]}
        \label{fig:agentic_memory_characteristics}
    \end{subfigure}
    \vspace{-15pt}
    \caption{Heterogeneous memory design opportunities and challenges for agentic LLM inference. No memory offers both the capacity and bandwidth that prefill and decode demand, so the memory system must be searched rather than picked.}
    \vspace{-15pt}
    \label{fig:combined_motivation}
\end{figure}

\footnotetext[1]{The workload trace is obtained on the LLaMA~3.3~70B model on H100.}

\footnotetext[2]{The bandwidth and capacity experiments are conducted using input/output sequence lengths of (1.4K, 0.2K) and (90K, 8K) on the LLaMA~3.3~70B model with the base configuration in \Cref{tab:sample_config} under weight-stationary execution.}

Large language models (LLMs) have fueled a surge in applications that require highly efficient inference systems. In particular, recent advances in agentic applications such as computer-use agents (CUAs)~\cite{osworld, agent_s2}, autonomous coding agents~\cite{longbench, ouyang2025kernelbenchllmswriteefficient, kernelcraft}, and web-use agents (WUAs)~\cite{webagent, osworld} introduce highly dynamic and unpredictable memory access patterns. Unlike conventional chatbot inference, where session lengths and sequence patterns are relatively predictable, agentic systems interact with external environments: this dynamic interaction leads to rapid context growth and extremely long traces from tool-calling and reasoning.

\begin{figure*}[h!]
    \centering
    \includegraphics[width=0.88\linewidth]{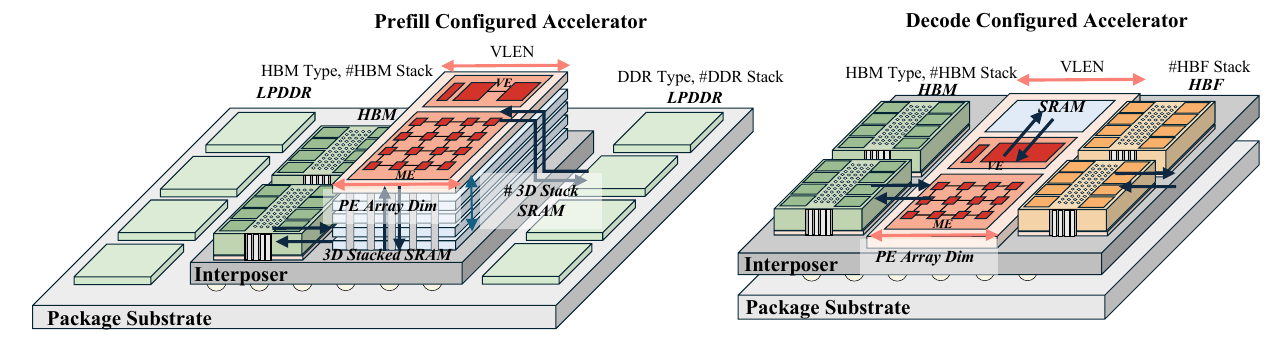}
    \caption{NPU designs specialized for each serving phase (i.e. Prefill and Decode). The black edges represent all data movements between the compute and memory. The parameters listed in the \Cref{tab:design_space} define the design space of the NPU system.}
    \label{fig:mem_sys}
\end{figure*}

\begin{table*}[t]
\centering
\caption{Comparison with existing NPU and system-level co-design works for LLM inference. MemExplorer considers configurable compute architectures, the broadest set of on-chip and off-chip memory devices, and data movement optimizations under both prefill--decode (PD) and attention--FFN (AF) disaggregation scenarios.}
\label{tab:related_work_comparison}
\small
\renewcommand{\arraystretch}{1.08}
\setlength{\tabcolsep}{4.5pt}
\begin{tabular}{lllccccccccc}
\toprule
\multirow{2}{*}{\textbf{Work}} &
\multirow{2}{*}{\textbf{Target Device}} &
\multirow{2}{*}{\textbf{CM}} &
\multicolumn{2}{c}{\textbf{On-chip}} &
\multicolumn{4}{c}{\textbf{Off-chip}} &
\multirow{2}{*}{\textbf{DM}} &
\multirow{2}{*}{\textbf{Disagg.}} &
\multirow{2}{*}{\textbf{Hetero.}} \\
\cmidrule(lr){4-5}\cmidrule(lr){6-9}
& & & \textbf{SRAM} & \textbf{3D-SRAM}$^{\dagger}$ & \textbf{HBM} & \textbf{LPDDR} & \textbf{GDDR} & \textbf{HBF}$^{\dagger}$ & & & \\
\midrule
PLENA~\cite{plena}              & NPU & \cmark & \cmark & \xmark & \cmark & \cmark & \xmark & \xmark & \cmark & --    & \xmark \\
Splitwise~\cite{Splitwise}      & GPU         & \xmark & \cmark & \xmark & \cmark & \xmark & \xmark & \xmark & \cmark & PD    & \xmark \\
LLMCompass~\cite{llmcompass}    & GPU         & \xmark & \cmark & \xmark & \cmark & \xmark & \cmark & \xmark & \cmark & PD    & \xmark \\
LLMServing2.0~\cite{llmserving2}    & GPU/NPU         & \xmark & \cmark & \xmark & \cmark & \cmark & \cmark & \xmark & \cmark & PD+AF  & \cmark \\
\midrule
\textbf{MemExplorer}            & NPU & \cmark & \cmark & \cmark & \cmark & \cmark & \cmark & \cmark & \cmark & PD+AF & \cmark \\
\bottomrule
\end{tabular}
\vspace{2pt}
\caption*{\footnotesize
\textbf{CM.} compute--memory co-design.
\textbf{DM} data-movement co-optimization: dataflow, storage, and bandwidth strategy.
\textbf{Disagg.} disaggregated system synthesis: which split strategies can be explored.
\textbf{Hetero.} heterogeneous cross-device specialization per disaggregated stage.
$^{\dagger}$Emerging memory technologies.}
\vspace{-15pt}
\end{table*}

In addition, agentic LLMs~\cite{kimi1m, glm4, qwen1m} now support context windows of millions of tokens, creating substantial memory-capacity demands; for some model and precision configurations, the KV cache alone can exceed 50 GB during agentic workloads. Concurrently, this dynamic interaction creates intense, rapidly changing memory pressure: sudden bursts of agentic interaction result in volatile memory access patterns and bandwidth requirements. These patterns expose a key limitation of today's accelerators: constrained by a single, homogeneous memory architecture (typically HBM), they struggle to simultaneously meet the dynamic bandwidth needs and varying memory capacity demands of long-context agentic workloads at different inference serving phases.

To address these limitations, emerging inference systems are shifting toward heterogeneous memory architectures across disaggregated accelerators.
The driving asymmetry is fundamental and not specific to agentic serving: prefill is compute- and throughput-bound, whereas decode is latency- and bandwidth-bound. Industry has begun to specialize the memory system to match each phase. NVIDIA's Vera Rubin platform~\cite{nvidia_vera_rubin_platform}, for instance, pairs Rubin GPUs backed by HBM4 (288 GB and up to 22 TB/s per node) for high-throughput prefill with Groq 3 LPX dataflow processors that aggregate 128 GB of on-chip SRAM, \emph{without} co-packaged DRAM, to reach 40 PB/s for latency-sensitive decode. Here, the GPU and LPX, built on different memory technologies, cooperate on a single serving task, signaling a broader shift away from uniform designs. AWS follows the same recipe, pairing Trainium with Cerebras WSE~\cite{cerebras_aws}. However, in each of these systems, the device pairing remains a \emph{fixed} point design: the mapping of memory technology to phase is hand-chosen and frozen, demonstrating the opportunity but not how to navigate it. Agentic workloads, with their long contexts and volatile memory demands, make getting this choice right more consequential, motivating a systematic way to explore the space rather than commit to a single configuration.

Emerging memory technologies open an opportunity here, but exploiting it systematically raises three challenges.
\textbf{Challenge 1: Rapidly integrating and evaluating emerging memory technologies.} As shown in \Cref{tab:offchip_memory_params}, memory technologies exhibit distinct trade-offs in bandwidth, capacity, latency, power, area, and cost. A new technology must therefore be efficiently modeled and evaluated alongside existing memories to determine its role in both the memory hierarchy within each device and the heterogeneous architecture across disaggregated devices.
\textbf{Challenge 2: these technologies cannot be evaluated in isolation.} Whether a 3D-stacked SRAM tier or an HBF stack pays off is not a property of the memory itself. It depends on the compute throughput that tier has to feed, how much of the working set it keeps on chip, and the data movement between the two. A memory model alone therefore cannot predict inference performance, and the technologies have to be selected jointly with the compute architecture and the execution strategy rather than dropped into a fixed accelerator.
\textbf{Challenge 3: disaggregated serving requires device-level specialization.} As shown in \Cref{fig:agentic_memory_characteristics}, the bandwidth and capacity requirements differ significantly between prefill and decode. This imbalance is reflected in the execution traces shown in \Cref{fig:motivation_agentic}. Although decode produces far fewer tokens than prefill processes, it can still account for a large fraction of the end-to-end latency. A single accelerator tuned for one stage is therefore compromised on the other, so the exploration must specialize devices across the system rather than optimize each one in isolation.

Given these challenges, our research question is: {\em How can we systematically explore this memory design space, jointly with compute and data movement, to identify optimal configurations of emerging memory technologies for next-generation agentic LLM inference systems with heterogeneous NPUs?}

\begin{table*}[h!]
\centering
\caption{Technology parameters of different memory devices. Values marked with $^{*}$ denote experiment data.
Values marked with $^{\dagger}$ denote values computed using the provided scaling factors.}
\label{tab:offchip_memory_params}
\setlength{\tabcolsep}{1.5pt}
\renewcommand{\arraystretch}{0.92}
\small
\begin{tabular}{lccccccc p{3.5cm}}
\toprule
Memory Type & Latency & Capacity & Bandwidth & Shoreline & $p_{\mathrm{bg}}$ (mW/GB) & $e_{\mathrm{read}}$ (pJ/bit)  & $e_{\mathrm{write}}$ (pJ/bit) & Note \\
\midrule
\rowcolor{gray!15}
\multicolumn{9}{c}{\textbf{On-Chip Memory}} \\
SRAM 1 Die & $\sim 1.5ns$ & 256 MB & 4TB/s & — & $\sim 10k- 50k^{*}$& $\sim 0.1^{*}$ & $\sim 0.1^{*}$& Obtained by Experiments \\
\midrule
\rowcolor{gray!15}
\multicolumn{9}{c}{\textbf{Off-Chip Memory}} \\
HBM3E 8H 1 Stack & $\sim100ns$ & 24 GB & 1 TB/s & $\sim11mm$ & $\sim50 - 100^{*}$ & $\sim 3^{*}$ & $\sim 3.6^{*}$ & Obtained by Experiments \\
HBM4 12H 1 Stack & $\sim100ns$ & 36 GB & 2 TB/s & $\sim15mm$ & $\sim50 - 100^{\dagger}$ & $\sim 2.2^{\dagger}$ & $\sim 2.4^{\dagger}$ & 40\% energy efficiency than HBM3E~\cite{patsnap_hbm_wars_2025}.\\
LPDDR5X 1 Pkg\cite{micron_lpddr5x} & $\sim50ns$  & 16 GB & 76.8 GB/s & $\sim4.1mm$ & $\sim7.65^{*}$ & $\sim 5^{*}$ & $\sim 6.5^{*}$ & Obtained by Experiments \\
LPDDR6 1 Pkg\cite{synopsys_lpddr6_vs_lpddr5x_2026} & $\sim50ns$  & 16 GB & 172.8 GB/s & $\sim4.5mm$ & $\sim6.12^{\dagger}$ & $\sim 3.75^{\dagger}$ & $\sim 4.87^{\dagger}$ & 20\% to 30\% more energy efficient than LPDDR5X \cite{moon_jedec_gddr7_2025}\\
GDDR6 1 Chip\cite{micron_gddr6_flyer} & $\sim12ns$ & 2 GB & 64 GB/s & $\sim11mm$ & $\sim100^{*}$ & $\sim 7^{*}$ & $\sim 8.8^{*}$ & Obtained by Experiments\\
GDDR7 1 Chip & $\sim12ns$ & 3 GB & 128 GB/s & $\sim11mm$ & $\sim120^{\dagger}$ & $\sim 5.6^{\dagger}$ & $\sim 7.0^{\dagger}$ & 20\% more energy efficient than GDDR6~\cite{techpowerup_gddr7_voltage_2023}.  \\
\midrule
\rowcolor{gray!15}
\multicolumn{9}{c}{\textbf{Emerging Memory Technologies}} \\
3D Stacked 1 Layer & $\sim 5ns$ & 1GB & 8TB/s & — & $\sim 10k- 50k^{*}$& $\sim 0.1^{*}$ & $\sim 0.1^{*}$& Obtained by Experiments \\
HBF 1 Stack\cite{kim_hbf_roadmap_2026,HBF} & $\sim1\mu s$ & 384 GB & 1 TB/s & $\sim8.25mm$ & $\sim300^{\dagger}$ & $\sim 6^{\dagger}$ & $\sim 10^{\dagger}$ & 4$\times$ $p_{\mathrm{bg}}$ and 2$\times$ $e_{\mathrm{read}}$ and $e_{\mathrm{write}}$ than HBM3E~\cite{HBF}.\\
\bottomrule
\end{tabular}
\end{table*}

We present \textbf{MemExplorer}, a systematic framework for synthesizing heterogeneous, disaggregated NPU systems. It searches the memory architecture, the compute architecture, and the software execution strategy as a single design space, instead of tuning one axis at a time. We will release MemExplorer as an open-source framework upon acceptance. Our key contributions are as follows, where the first three contributions address the three challenges above, respectively:

\begin{itemize}[leftmargin=1em]
\item
\textbf{Broad coverage of emerging memory technologies.}
MemExplorer brings emerging memory technologies, namely 3D-stacked SRAM and HBF,
into the design space alongside conventional DRAM. It models every technology
under a unified set of capacity, bandwidth, latency, power, and shoreline-area
constraints (\Cref{tab:offchip_memory_params}), allowing any technology to occupy
any tier of a multi-level hierarchy and making both the memory hierarchy and
cross-device heterogeneity searchable dimensions. On agentic workloads,
including these technologies in the search improves TPS by
\textbf{1.74$\times$} and Token/J by \textbf{1.71$\times$} over the same joint
search restricted to commercially available memory technologies.

\item
\textbf{Joint next-gen memory architecture synthesis with compute.} MemExplorer
co-synthesizes the compute architecture, the memory hierarchy, and the
data-movement strategy as one design space, rather than tuning memory in
isolation as the works in \Cref{tab:related_work_comparison} do. The
data-movement strategy spans dataflow, on-chip storage priority, and off-chip
bandwidth allocation, so the search captures the balance between compute
throughput, on-chip residency, and off-chip traffic that actually sets
end-to-end performance.

\item
\textbf{Systematic cross-device specialization.} MemExplorer specializes each device for the phase it serves in a system, under both prefill--decode and attention--FFN disaggregation. This admits configurations that look unbalanced in isolation yet win at the system level, which device-local optimization cannot discover.

\item
\textbf{New architectural insights.} MemExplorer enables phase-specific memory selection and identifies the most suitable technology for each serving phase. Prefill benefits most from large, high-bandwidth 3D-stacked on-chip SRAM: under the same power budget, the optimized prefill chip reaches \textbf{2.3$\times$} and \textbf{3.23$\times$} higher Token/J than the baseline NPU and the H100, respectively. Decode instead favors larger-capacity, lower-bandwidth off-chip memory, delivering \textbf{1.93$\times$} and \textbf{2.72$\times$} higher Token/J than the baseline NPU and the H100.
\end{itemize}

\section{Unified Memory System Modeling}

We establish an analytical memory model with two objectives:
(1)~characterizing the memory behavior intrinsic to agentic LLM inference for further integration into the co-design framework; and
(2)~remaining extensible to emerging memory technologies so that the framework retains its utility as the
technology landscape evolves.
Achieving both objectives simultaneously is non-trivial. We therefore abstract each memory technology using the smallest set of parameters required to evaluate its impact on NPU-level co-design. Our key observation is that, despite their diversity, memory technologies relevant to NPU co-design can be faithfully compared
along a compact parameter set spanning physical layers (memory stack count) and performance (latency, capacity, bandwidth, energy), as consolidated in the header of \Cref{tab:offchip_memory_params}.
In the following subsections, we first detail how these parameters are instantiated for established technologies and the constraints governing their integration. We then present the hierarchical analytical model that underpins the co-design framework. Finally, we demonstrate extensibility through two case studies on 3D-stacked SRAM and High Bandwidth Flash.

\subsection{Established Memory Technologies}

In this work, we consider on-chip SRAM and three widely deployed off-chip memory families: GDDR, LPDDR, and HBM. Both on-chip and off-chip memory parameters are summarized in \Cref{tab:offchip_memory_params}.

\paragraph{Memory Stack Count.}
This constraint applies exclusively to off-chip memory; on-chip SRAM is fabricated directly on, or bonded atop, the compute die and therefore does not consume die-shoreline resources. For off-chip technologies, memory devices are physically placed along selected edges of the compute die and connected through high-speed PHY interfaces, collectively referred to as the die shoreline~\cite{shorline}. This imposes a fundamental physical bound on the number of memory stacks that can be integrated around a single compute die. In modern accelerator designs, only a subset of the shoreline (typically two edges) is allocated for memory attachment, while the remaining edges are reserved for PCIe, scale-up interconnects (e.g., NVLink), and other I/O. Each off-chip stack occupies $L_{\mathrm{PHY}} \approx 4$--$15\,$mm of shoreline depending on the technology (see \Cref{tab:offchip_memory_params}), bounding the maximum number of attachable stacks as

\begin{equation}
N_{\mathrm{stack}} \;\lesssim\;
\left\lfloor
\frac{L_{\mathrm{mem}}}{L_{\mathrm{PHY}} + L_{\mathrm{margin}}}
\right\rfloor
\label{eq:memory_stack_limit}
\end{equation}
where $L_{\mathrm{mem}}$ is the die-edge length reserved for memory. This constraint is further tightened by the lithography reticle limit: current DUV/EUV steppers impose a maximum die exposure field of approximately $26\,\mathrm{mm} \times 33\,\mathrm{mm}$~\cite{asml,waferscale}, giving an upper bound of $L_{\mathrm{mem}} \le 2 \times 33\,\mathrm{mm}$. Consequently, even under an unconstrained power budget, the number of off-chip stacks is hard-bounded by physical die geometry~\cite{muchisim}.

\paragraph{Performance Indicators.}
We characterize each technology along three performance axes: I/O access latency (Latency), per-die capacity (Capacity), and peak bandwidth (Bandwidth). These metrics form the standard basis for memory system evaluation in prior accelerator studies~\cite{memory_metric_basic} and are directly obtainable from vendor datasheets and JEDEC specifications. Together, they determine the rate at which data can be supplied to the compute units and the volume of model state that can reside at each hierarchy level, both of which critically govern LLM inference throughput.

In addition, we decompose memory energy into three components: static background power ($p_{\mathrm{bg}}$), per-bit read energy ($e_{\mathrm{read}}$), and per-bit write energy ($e_{\mathrm{write}}$). This decomposition is sufficient for our transactional memory model because LLM inference memory accesses are dominated by large, contiguous bulk transfers (each modeled as a single read or write transaction), rather than fine-grained random accesses that would require row-buffer or bank-conflict modeling. Background power captures the capacity-dependent leakage cost that persists regardless of access activity, while the per-bit dynamic terms capture the access-dependent cost that scales with bandwidth utilization. Together, these three coefficients parameterize the power equation in \Cref{eq:hbm_power_model}; a discussion of the memory power model is presented in \Cref{sec:system_model}.

\subsection{Proposed Analytical Model for Memory Hierarchy}
\label{sec:memory_traffic_model}

LLM inference primarily streams large tensors from memory, so we model each tensor read or write as a single memory operation. For example, loading the K or V cache for an attention layer is one operation, with its bandwidth demand determined by the compute units.

For on-chip memory, we model bandwidth utilization for matrix and vector operations separately because they exhibit distinct access patterns. For off-chip memory, we model an $L$-level hierarchy composed of multiple memory technologies, where level $0$ is the compute unit and increasing indices denote deeper memory levels. Data moves toward compute across adjacent levels, from level $i$ to level $i-1$.

For each memory level $i$, we define the following quantities:
\begin{itemize}[leftmargin=1em]

\item $B_i^{\mathrm{peak}}$ and $B_i^{\mathrm{eff}}$: peak and effective bandwidth, respectively, across the boundary from level $i$ to level $i-1$.
\item $\lambda_i$: fixed access/setup latency for transferring data from level $i$ to level $i-1$.
\item $x$: total amount of data requested by the memory operation.
\item $x_i$: total amount of data that must be delivered across the boundary from level $i$ to level $i-1$.
\item $\alpha_i$: local resident fraction at level $i$, i.e., the fraction of $x_i$ that is already resident at level $i$ and can be directly served from this level.
\end{itemize}

Thus, the amount of data locally resident at level $i$ is
\begin{equation}
    x_i^{\mathrm{local}} = \alpha_i x_i\,,
\end{equation}
and the missing amount that must be refilled from deeper levels is
\begin{equation}
    x_i^{\mathrm{refill}} = (1-\alpha_i)x_i\,.
\end{equation}

When level $i$ simultaneously serves level $i-1$ and is refilled from level $i+1$, the upward transfer and the refill traffic may share the bandwidth resources of level $i$. We therefore model the effective bandwidth available to the upward transfer as
\begin{equation}
    B_i^{\mathrm{eff}} = \max\left(0, B_i^{\mathrm{peak}} - B_{i+1}^{\mathrm{refill}}\right)\,.
\end{equation}
If no simultaneous refill occurs, or if the upward-transfer path and the refill path do not share bandwidth resources, then $B_{i+1}^{\mathrm{refill}}=0$ and $B_i^{\mathrm{eff}}=B_i^{\mathrm{peak}}$.

The latency for transferring the requested data across the boundary from level $i$ to level $i-1$ is
\begin{equation}
    \tau_i(x_i) = \lambda_i + \frac{x_i}{B_i^{\mathrm{eff}}}
\end{equation}
where $x_i$ is the total amount of data transferred across this boundary. It consists of the locally resident portion $\alpha_i x_i$ and the portion refilled from deeper levels, $(1-\alpha_i)x_i$.

\paragraph{Double-Buffering Transfer Model.}

With double buffering, the transfer from level $i$ to level $i-1$ can overlap with the refill from level $i+1$ to level $i$. Therefore, the effective latency is determined by the slower of the current-level upward transfer and the deeper-level refill:
\begin{equation}
    T_i(x_i,[\alpha_i,\ldots,\alpha_L]) = \max \left( \tau_i(x_i), T_{i+1}(x_{i+1},[\alpha_{i+1},\ldots,\alpha_L]) \right)
\label{eq:recursive_transfer}
\end{equation}
where $x_{i+1}=(1-\alpha_i)x_i$ denotes the amount of data that is not locally resident at level $i$ and must therefore be refilled from the next deeper level.

This formulation captures the overlap enabled by double buffering. If the current-level upward transfer is slower, i.e., $\tau_i(x_i) \geq T_{i+1}$, the deeper-level refill is fully hidden. Otherwise, if $T_{i+1} > \tau_i(x_i)$, the refill from the deeper memory level becomes the bottleneck. When overlap is not supported, the model can be reduced to a serialized form by replacing the maximum in \Cref{eq:recursive_transfer} with a summation:
\begin{equation}
    T_i(x_i,[\alpha_i,\ldots,\alpha_L]) = \tau_i(x_i) + T_{i+1}(x_{i+1},[\alpha_{i+1},\ldots,\alpha_L])\,.
\end{equation}

\subsection{Emerging Memory Technologies}
Beyond established devices, two emerging technologies, 3D-Stacked on-chip SRAM and High Bandwidth Flash (HBF). These two technologies occupy qualitatively new positions in the design space that existing models fail to capture. Their combined parameters are reflected in~\Cref{tab:offchip_memory_params}.

\paragraph{3D-Stacked SRAM}
Conventional 2D SRAM faces increasing area and leakage constraints at advanced process nodes~\cite{yu2022semiconductor}. By vertically stacking SRAM dies, 3D-stacked SRAM provides substantially greater on-chip capacity and bandwidth while retaining low latency and energy, as summarized in \Cref{tab:offchip_memory_params}. Its main limitation is the increased power density and thermal cost of additional layers. This makes it attractive for holding active KV-cache data on-chip and reducing accesses to off-chip memory.

\paragraph{High Bandwidth Flash (HBF)}
HBF~\cite{kim_hbf_roadmap_2026,HBF} combines vertically stacked NAND Flash, a DRAM buffer, and a high-bandwidth interface to provide substantially higher capacity density than HBM. However, this capacity comes with significant power and energy penalties. Its static background power is approximately $4\times$ higher than that of HBM~\cite{HBF,emergentmind_hbf_2026}, while its per-bit read and write energies are roughly $2\times$ higher~\cite{ma2026llmhardware}. HBF is therefore unsuitable as the primary bandwidth tier, but is well suited to high-capacity weight storage and KV-cache overflow in long-context workloads.

\section{Memory System Exploration for Heterogeneous NPUs}
During LLM inference, the prefill and decode stages exhibit different compute and memory access patterns. Prefill processes long input sequences with large activation footprints, while decode generates tokens incrementally and is dominated by KV-cache access and reuse. These differences make it challenging for a single NPU to simultaneously optimize both phases, motivating heterogeneous memory system design and the potential for Prefill–Decode disaggregation~\cite{distserve}.

Prior work, such as PLENA~\cite{plena} and RPU~\cite{rpu}, has analyzed the computation characteristics of the prefill and decode stages. However, these works assume fixed memory configurations, which limits their ability to capture phase-specific memory demands. In this work, we further investigate the memory behavior of these stages and identify distinct bandwidth and capacity requirements across the two phases.

\paragraph{Systematic Memory Bandwidth Requirement}
During the prefill stage, especially when activations exceed on-chip capacity, both activations and weights must be loaded simultaneously from off-chip memory, resulting in high bandwidth demand. In contrast, during the decode stage, only weights and KV cache are streamed from memory, leading to lower overall bandwidth requirements.

\paragraph{Systematic Memory Capacity Requirement}
Memory capacity plays a more critical role during the decode stage. As observed in prior work~\cite{plena}, decode often suffers from low compute utilization when the batch size is limited by memory capacity. This effect is further amplified in agentic workloads, where long contexts and large KV caches significantly increase memory requirements. When capacity is insufficient, batch size must be reduced, resulting in smaller workloads and poor systolic array utilization. Increasing memory capacity enables larger batch sizes, thereby improving compute utilization and throughput. In contrast, the prefill stage is typically less sensitive to capacity constraints, as it is primarily compute-bound, with compute units already operating at high utilization.

These observations highlight that prefill and decode benefit from different memory configurations: prefill favors high-bandwidth memory to sustain compute throughput, whereas decode benefits from larger-capacity tiers to improve utilization and efficiency, motivating heterogeneous NPU designs.

\section{Algorithm-aware Systematic Co-design}

MemExplorer enables systematic co-design of the memory hierarchy, compute architecture, and software strategies in a holistic manner for heterogeneous NPU design. The framework consists of the following key components, which are described in the subsequent sections.

\begin{enumerate}[label=\alph*), leftmargin=1em]
    \item \textbf{System Model} includes both memory and compute models. It estimates system performance and power.

    \item \textbf{Data Movement Model} captures dataflow, data placement, and bandwidth allocation across memory levels to account for workload-specific execution strategies.

    \item \textbf{Workload Serving Engine} takes model configurations and workload traces (including prompt and generated token counts) as input, and models the prefill, decode-only, and combined performance using the system model.

    \item \textbf{Co-Design Exploration} jointly explores the NPU architecture, heterogeneous memory hierarchy, and software strategies, enabling systematic design space exploration (DSE) to optimize performance and power efficiency.
\end{enumerate}

\begin{figure}
    \centering
    \includegraphics[width=\linewidth]{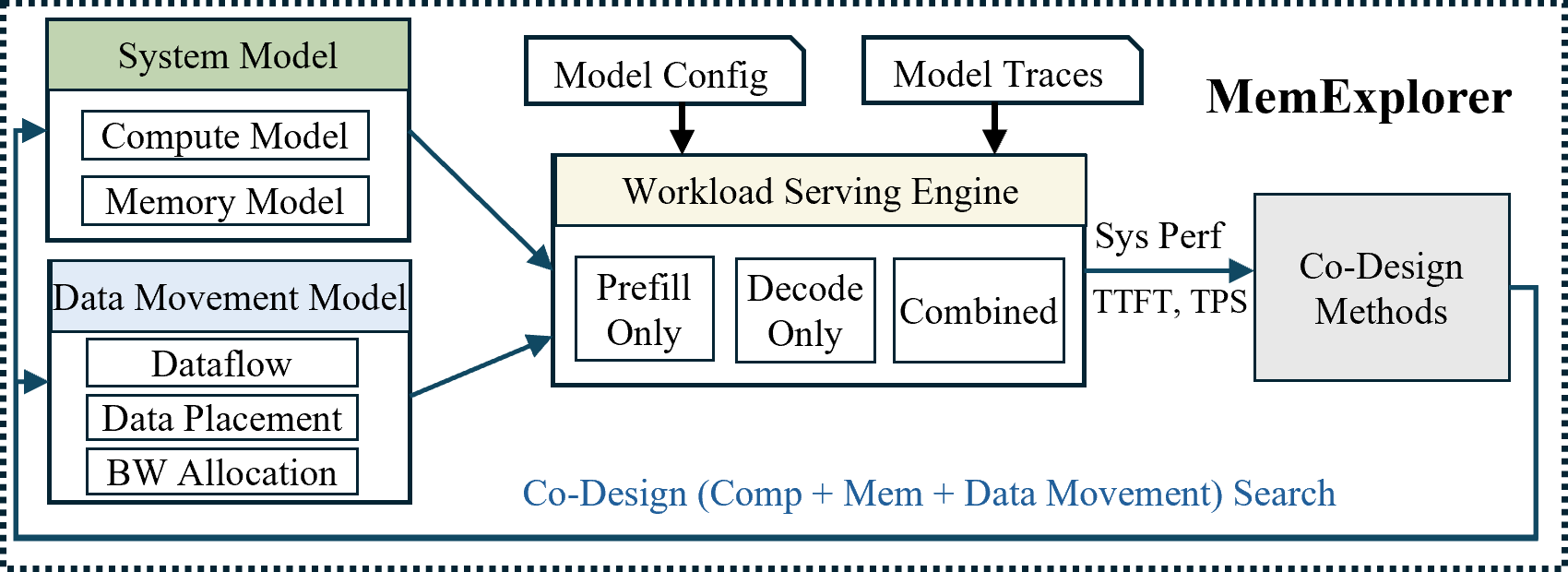}
    \caption{Co-design workflow of the MemExplorer framework. Given a Hugging Face model configuration and workload traces, MemExplorer iteratively optimizes the NPU system across compute, memory, and data movement.}
    \label{fig:Mem_Explorer}
\end{figure}

\subsection{System Model}
\label{sec:system_model}
To obtain the overall performance model, we integrate the analytical hierarchical memory model (discussed in \Cref{sec:memory_traffic_model}) with the PLENA~\cite{plena} performance analytical model. The memory model characterizes data movement across on-chip and multi-level off-chip memory, including bandwidth utilization, latency, and power consumption. The compute analytical model is based on PLENA~\cite{plena}, which is highly configurable in terms of compute dimensions and memory controller bandwidth.

The total system power consumption is primarily driven by two components: (1) the accelerator compute unit executing operations, and (2) data movement across the heterogeneous memory hierarchy. Modeling both components comprehensively is essential for accurate system-level evaluation.
\begin{itemize}[leftmargin=1em]
\item \textbf{Compute Power Modeling}
We construct an analytic power model by decomposing the accelerator into key components, such as the systolic array matrix units. For each component, we generate representative design samples and obtain power estimates through synthesis using Synopsys Design Compiler with the 7\,nm OpenROAD predictive PDK~\cite{openroad}. We then fit parametric models to these samples to capture the power behavior of individual components. Finally, we aggregate these component-level models to obtain an overall power estimation for PLENA.

\item \textbf{Memory Power Modeling} The memory power model is estimated by aggregating the power consumption of individual memory units. The model captures both read and write behaviors to provide accurate power estimation. The formulation is shown in \Cref{eq:hbm_power_model}, and the coefficients are summarized in \Cref{tab:offchip_memory_params}. Bandwidth utilization is derived from the memory traffic model for each memory unit.

\end{itemize}

\begin{equation}
p(c, b_{\mathrm{read}}, b_{\mathrm{write}})
=
p_{\mathrm{bg}} c
+ e_{\mathrm{read}} b_{\mathrm{read}}
+ e_{\mathrm{write}} b_{\mathrm{write}}
\label{eq:hbm_power_model}
\end{equation}

where:
\begin{itemize}
    \item $c$ is the total memory capacity (GB),
    \item $b_{\mathrm{read}}$ is the read bandwidth (bit/s), and
    \item $b_{\mathrm{write}}$ is the write bandwidth (bit/s).
\end{itemize}

\begin{figure}[htbp]
    \centering
    \includegraphics[width=0.9\linewidth]{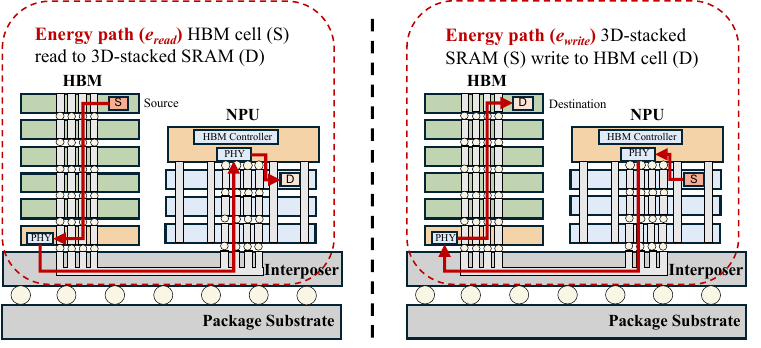}
    \caption{\textbf{Memory Read/Write Power Measurement}. The path for $e_{\mathrm{read}}$ (left) outlines the end-to-end energy consumed when retrieving data from the memory cells and transferring it to the PLENA compute die. Conversely, $e_{\mathrm{write}}$ (right) accounts for the energy required to move data from the PLENA compute die and write it into the memory cells.}
    \label{fig:power_model}
\end{figure}

\subsection{Data Movement Model}
\label{sec:software_strategies}

As shown in \Cref{tab:design_space}, the data movement model supports the simulation of different inference strategies, which are discussed below.

\begin{itemize}[leftmargin=1em]
    \item \textbf{Dataflow Strategy}: This strategy determines how weights, inputs, and outputs are accessed during computation (e.g., weight-stationary, input-stationary, or output-stationary)\cite{dataflow}. When one data type is stationary, it remains on-chip while the remaining data are streamed from memory. For large off-chip data, frequent memory accesses can significantly degrade performance, making dataflow selection critical for efficient execution.
    \item \textbf{On-Chip Storage Priority}: This strategy abstracts compiler-level optimizations by prioritizing which data types (weights, activations, KV cache, or intermediate results) are stored on-chip. Since most NPUs operate on a limited set of data types, this higher-level abstraction enables efficient modeling of storage decisions without requiring detailed compiler optimizations.
    \item \textbf{Off-Chip Bandwidth Priority}: Most NPUs support concurrent streaming of multiple data types using separate buffers, but limited off-chip bandwidth may prevent simultaneous saturation. Therefore, bandwidth allocation priority must be defined. To limit design space complexity, we adopt a fixed allocation policy, assigning 75\% bandwidth to matrix data and 25\% to vector data when matrix priority is selected.

\end{itemize}

\subsection{Workload Serving Engine}

To evaluate prefill-only and decode-only throughput for a specific workload, we model each stage independently by identifying the cases for maximum throughput.

\begin{itemize}[leftmargin=1em]
    \item \textbf{Prefill Throughput}
    During the prefill stage, throughput is primarily limited by compute and memory bandwidth when processing long input sequences. Hence, we are simply treating the single batch for evaluation.

    \item \textbf{Decode Throughput}
    During the decode stage, throughput is dominated by KV-cache growth and batch-level parallelism. Hence, we need to set the batch size to the max. The maximum batch size is then derived based on the available memory capacity.  The memory footprint is computed by considering model weights and KV cache, which scale with batch size, input tokens, and generated tokens.
\end{itemize}

\subsection{Co-Design Exploration}

The architectural design space of an NPU design is inherently vast,
spanning compute array dimensions, multi-tier memory configurations
(including 3D-stacked SRAM, HBM, and other off-chip technologies),
data movement strategies,
as summarized in Table~\ref{tab:design_space}.
The cross-product of these parameters yields a design space
on the order of $10^{6}$ configurations,
making exhaustive evaluation prohibitively expensive.

We formulate the search as a multi-objective design-space exploration problem
that maximizes inference throughput while minimizing power consumption, subject
to a total system power budget and memory-package constraints, including the
area/shoreline limit in \Cref{eq:memory_stack_limit}. Each candidate design is
evaluated using the analytical performance and power models described above.

To approximate the Pareto frontier under a limited evaluation budget,
we employ Multi-Objective Bayesian Optimization (MOBO)~\cite{daulton2020differentiable}.
We first sample $N_{\mathrm{init}}{=}20$ initial configurations using Sobol
sequences, and then iteratively select new candidates by fitting Gaussian-process
surrogates to the observed designs and maximizing the Expected Hypervolume
Improvement (EHVI) acquisition function~\cite{emmerich2006single,daulton2020differentiable}.
The search terminates after $N_{\mathrm{total}}{=}100$ evaluations, yielding the
Pareto-optimal designs used in our analysis.

\begin{table}[t]
\centering
\caption{Design space including compute, memory configurations, and data movement strategies.
On-chip storage represents allocating additional on-chip memory capacity to store frequently reused data (e.g., weights, activations, or KV cache) to reduce off-chip traffic.
Dataflow strategy refers to different GEMM execution strategies, including weight stationary (WS), input stationary (IS), and output stationary (OS), which determine data reuse patterns across the memory hierarchy.
Bandwidth allocation priority represents redistributing limited off-chip memory bandwidth across different ports to maximize utilization under bandwidth constraints.}

\label{tab:design_space}
\small
\setlength{\tabcolsep}{2pt}
\renewcommand{\arraystretch}{1.15}
\begin{tabularx}{\columnwidth}{p{3.1cm} X}
\toprule
\textbf{Parameter} & \textbf{Design Space} \\
\midrule

\rowcolor{gray!15}
\multicolumn{2}{c}{\textbf{Compute Configuration}} \\
PE Array Dim & \(\{128\times128,\;64\times256,\;32\times512,\;16\times1024\}\) \\
VLEN & \(\{128,\;256,\;512,\;1024,\;2048\}\) \\

\midrule
\rowcolor{gray!15}
\multicolumn{2}{c}{\textbf{On-chip Memory Configuration}} \\
3D-Stacked SRAM & \(\{0,\;1,\;2,\;3,\;4\}\) \\
Conventional SRAM & \(\{\text{Enabled},\;\text{Disabled}\}\) \\

\midrule
\rowcolor{gray!15}
\multicolumn{2}{c}{\textbf{Off-chip Memory Configuration}} \\

HBM & \makecell[l]{
Type \(\in\{\)
HBM3E (24GB, 1TB/s), \\
HBM4 (36GB, 2TB/s) \(\}\) \\
Stacks \(\in\{1,2,4,8\}\)} \\

\midrule
HBF & \makecell[l]{
HBF (384GB, 1TB/s) \\
Stacks \(\in\{1,2,4,8\}\)} \\

\midrule
GDDR & \makecell[l]{
Type \(\in\{\)
GDDR6 (2GB, 64GB/s), \\
GDDR7 (3GB, 128GB/s) \(\}\) \\
Stacks \(\in\{1,2,4,8\}\)} \\

\midrule
LPDDR & \makecell[l]{
Type \(\in\{\)
LPDDR5X (16GB, 76.8GB/s), \\
LPDDR6 (16GB, 172.8GB/s) \(\}\) \\
Stacks \(\in\{1,2,4,8\}\)} \\

\midrule
\rowcolor{gray!15}
\multicolumn{2}{c}{\textbf{Data Movement Strategy}} \\
Storage Priority (SP) & \(\{\text{Activation, KV Cache, Weight, Equal}\}\) \\
Dataflow Strategy (DS) & \(\{\text{WS, OS, IS}\}\) \\
BW Priority (BP) & \(\{\text{Matrix, Vector, Equal}\}\) \\

\bottomrule
\end{tabularx}
\end{table}

\section{Evaluation}
\label{sec: eval}

We evaluate MemExplorer by first conducting ablation studies on compute, data movement strategies, and memory hierarchy configurations to justify the co-design space, then performing end-to-end design space exploration under realistic power constraints to compare against GPU baselines on agentic workloads and emerging model architectures, and finally validating our analytical model against cycle-accurate emulation, RTL synthesis, and real GPU measurements.

\subsection{Experiment Setup}
\paragraph{Models and Datasets}
We evaluate several popular open-source Transformer-based large language models (LLMs), including
\llamaIIIPLUSBIG~\cite{llama3}, Qwen3~\cite{qwen3}, and LLaDA~\cite{llada}.
For agentic workloads, we focus on the BFCL~\cite{bfcl} and OSWorld~\cite{osworld} benchmarks.

To make the tracing methodology explicit, we collect agentic traces from two harnesses:
BFCL-Web Search Base~\cite{bfcl}, using the \texttt{web\_search\_base} tool-use task, and
OSWorld LibreOffice~\cite{osworld}, using the Calc task under the accessibility-tree setting.
These traces capture the prompt/generation-token distribution induced by each
agentic workflow. We run both workloads with \llamaIIIPLUSBIG using vLLM on an
NVIDIA H100 and use the maximum observed prompt/generation lengths in our
experiments: 114K/5K tokens for BFCL-Web Search Base and 90K/8K tokens for
OSWorld LibreOffice. We also include GSM8K~\cite{gsm8k} as a traditional chatbot
baseline with a representative 1.4K/0.2K prompt/generation-token setting.

\paragraph{Inference Evaluation}
We compare end-to-end system performance against GPU baselines, including the NVIDIA A100 (80\,GB SXM) and H100 (80\,GB SXM). The GPU experiments are conducted on a system running Ubuntu 22.04, CUDA 12.8, Python 3.11, PyTorch 2.8.0, and vLLM 0.10.

\paragraph{Precision Settings}

We set the activations, weights, and KV cache to the MX-INT data type with 4-bit precision, a block size of 16, and an 8-bit scale, following PLENA's balanced-precision configuration~\cite{plena}. This setting enables aggressive quantization while maintaining strong performance across a range of agentic tasks, including BFCL~\cite{bfcl}. All subsequent experiments use this configuration.

\subsection{Ablation Study}
To demonstrate the benefit of heterogeneous co-design across software, compute, and memory configurations, we compare our approach with three single-axis baselines that search software, compute, or memory alone. To separate the contribution of the search itself from that of the newly modelled technologies, we add a fourth baseline, Co-Design (W/O EM*), which searches all three axes jointly but restricts the memory hierarchy to conventional HBM and LPDDR tiers, excluding the emerging technologies (3D-stacked SRAM and HBF) that MemExplorer models. We report the performance of designs under PD disaggregation on both the OSWorld~\cite{osworld} agentic workload and the GSM8K~\cite{gsm8k} chatbot workload.

\begin{figure}[t]
    \centering
    \includegraphics[width=1\linewidth]{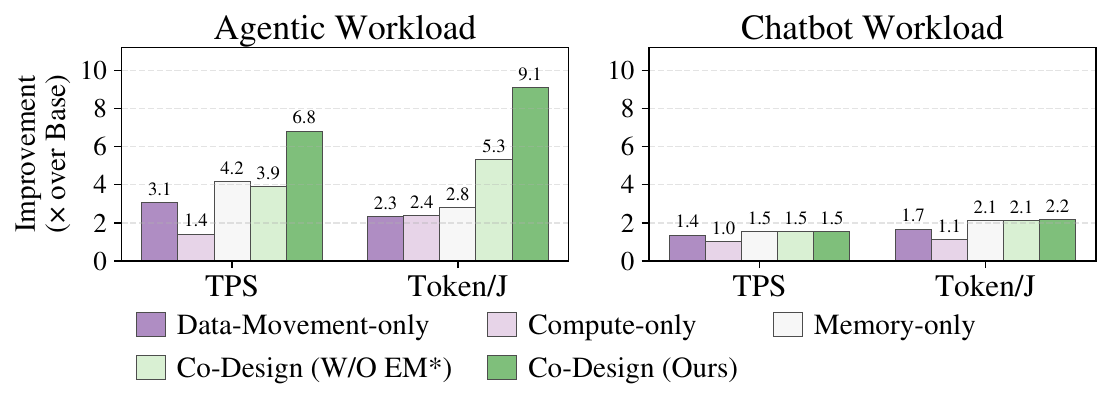}
    \caption{Joint data movement, compute, and memory co-design outperforms single-axis search under the same PD disaggregation setting, with larger gains on agentic workloads: up to 6.8$\times$ higher TPS and 9.1$\times$ Token/J over Base. Co-Design (W/O EM*) denotes the emerging memory technologies modelled by MemExplorer (3D-stacked SRAM and HBF);}
    \label{fig:codesign_improvement}
    \vspace{-5pt}
\end{figure}

As shown in \Cref{fig:codesign_improvement}, softwar This comparison demonstrates the benefits of co-design and quantifies the substantial performance gains enabled by emerging memory technologies.

\subsection{Device Specialization under PD Disaggregation}
\label{sec: co-optimize-hetero-sys}
Building on the ablations above, we now co-optimize a complete heterogeneous NPU system under PD disaggregation, jointly searching compute, memory, and data movement configurations for each phase. We explore the design space to identify Pareto-optimal configurations for two representative agentic language model workloads: BFCL-Web Search Base~\cite{bfcl} and OSWorld LibreOffice (OSWorld-L)~\cite{osworld}. During the search, we constrain the total Thermal Design Power (TDP) to 700\,W, matching the TDP of the NVIDIA H100 SXM equipped with 80\,GB of HBM3 memory. We also enforce the package shoreline constraint in \Cref{eq:memory_stack_limit}, which limits the number of memory stacks according to the die-edge length available for memory interfaces. The resulting Pareto-optimal configurations are summarized in \Cref{tab:sample_config}.

\begin{table*}[t]
\centering
\captionsetup{skip=2pt}
\caption{Pareto frontier samples selected from DSE for both prefill and decode optimization on the OSWorld task (input tokens: 90K, output tokens: 8K). TPS and Token/J are normalized to the Base configuration of each stage, with Token/J obtained from the measured TPS and average power. Note that for the 3D-stacked configuration, the compute layer is stacked on the SRAM and its area is excluded from the reported on-chip area, which is therefore smaller.}

\label{tab:sample_config}
\setlength{\tabcolsep}{1.15pt}
\renewcommand{\arraystretch}{0.95}
\small
\begin{tabular}{lcc cccc ccc c c ccc}
\toprule
\rowcolor{gray!15}
\multicolumn{15}{c}{\textbf{Prefill Optimization}} \\
\midrule

& \multicolumn{2}{c}{\textbf{Compute}}
& \multicolumn{4}{c}{\textbf{Memory Hierarchy}}
& \multicolumn{3}{c}{\textbf{Data Movement}}
& \multicolumn{1}{c}{\textbf{Power}}
& \multicolumn{1}{c}{\textbf{Area}}
& \multicolumn{3}{c}{\textbf{Perf}} \\
\cmidrule(lr){2-3}
\cmidrule(lr){4-7}
\cmidrule(lr){8-10}
\cmidrule(lr){11-11}
\cmidrule(lr){12-12}
\cmidrule(lr){13-15}

\textbf{Config}
& \textbf{PE Array} & \textbf{VLEN}
& \textbf{On-chip} & \textbf{L1 Off-chip} & \textbf{L2 Off-chip} & \textbf{L3 Off-chip}
& \textbf{SP} & \textbf{DS} & \textbf{BP}
& \textbf{TDP}
& \textbf{mm$^2$}
& \textbf{Batch} & \textbf{TPS} & \textbf{Token/J} \\

\midrule

Base & 2048x128 & 2048 & SRAM x1 & HBM3E x4 & None & None & Equal & OS & Equal & 300.1 & 252.7 & 16 & 1.00x & 1.00x \\

P1 & 2048x256 & 2048 & 3D Stacked x3 & HBM4 x2 & HBF x1 & None & Act & WS & Matrix & 697.1 & 197.3 & 16 & \best{6.71x} & 2.61x \\

P2 & 1024x512 & 2048 & 3D Stacked x2 & HBM4 x2 & LPDDR5X x8 & LPDDR5X x8 & Equal & WS & Equal & 570.0 & 190.1 & 16 & 4.93x & \best{2.68x} \\

\midrule

\rowcolor{gray!15}
\multicolumn{15}{c}{\textbf{Decode Optimization}} \\
\midrule

& \multicolumn{2}{c}{\textbf{Compute}}
& \multicolumn{4}{c}{\textbf{Memory Hierarchy}}
& \multicolumn{3}{c}{\textbf{Data Movement}}
& \multicolumn{1}{c}{\textbf{Power}}
& \multicolumn{1}{c}{\textbf{Area}}
& \multicolumn{3}{c}{\textbf{Perf}} \\
\cmidrule(lr){2-3}
\cmidrule(lr){4-7}
\cmidrule(lr){8-10}
\cmidrule(lr){11-11}
\cmidrule(lr){12-12}
\cmidrule(lr){13-15}

\textbf{Config}
& \textbf{PE Array} & \textbf{VLEN}
& \textbf{On-chip} & \textbf{L1 Off-chip} & \textbf{L2 Off-chip} & \textbf{L3 Off-chip}
& \textbf{SP} & \textbf{DS} & \textbf{BP}
& \textbf{TDP}
& \textbf{mm$^2$}
& \textbf{Batch} & \textbf{TPS} & \textbf{Token/J} \\

\midrule

Base & 2048x128 & 2048 & SRAM x1 & HBM3E x4 & None & None & Equal & OS & Equal & 300.1 & 252.7 & 16 & 1.00x & 1.00x \\

D1 & 1024x64 & 1024 & 3D Stacked x1 & HBM4 x2 & HBF x2 & LPDDR5X x8 & Act & WS & Matrix & 450.95 & 98.6 & 32 & \best{2.19x} & 1.28x \\

D2 & 2048x64 & 1024 & SRAM x1 & HBM3E x2 & HBF x1 & None & Act & WS & Matrix & 249.90 & 167.8 & 16 & 1.44x & \best{1.55x} \\

\bottomrule
\end{tabular}
\end{table*}

\begin{figure}[t]
    \centering
    \includegraphics[width=\linewidth]{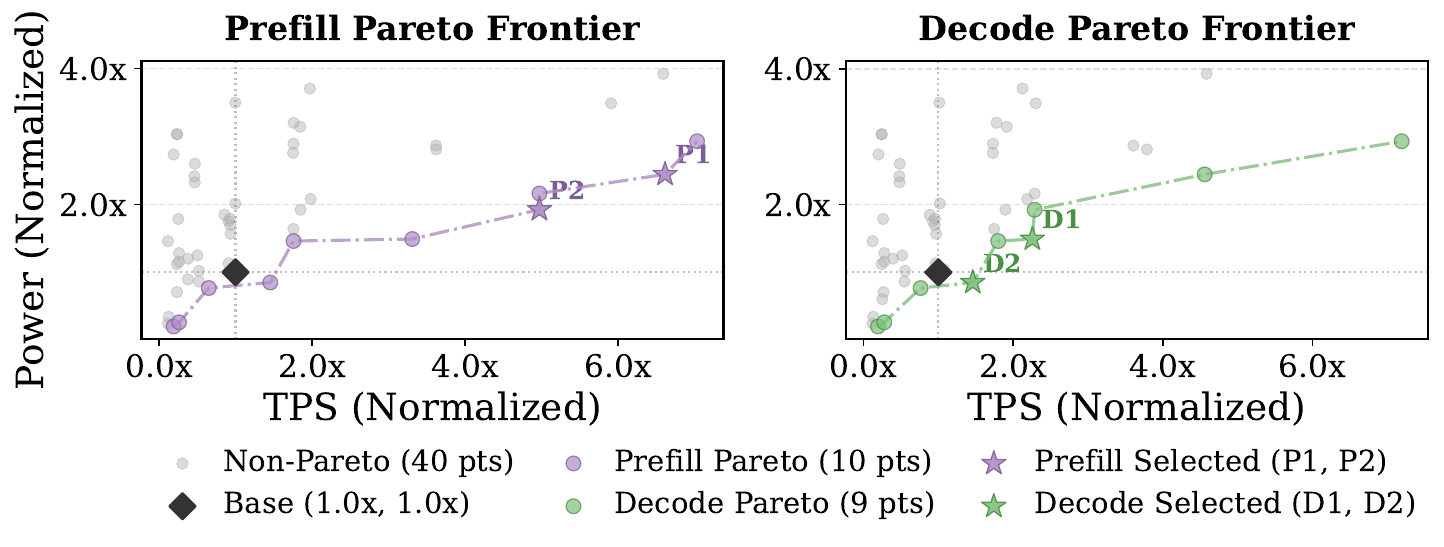}
    \caption{Representative designs selected from the Pareto frontier under the 700 W power constraint. P1 and D1 maximize TPS, whereas P2 and D2 maximize Token/J.}
    \label{fig:pareto_frontier}
\end{figure}

\begin{figure}[t]
    \centering
    \includegraphics[width=1\linewidth]{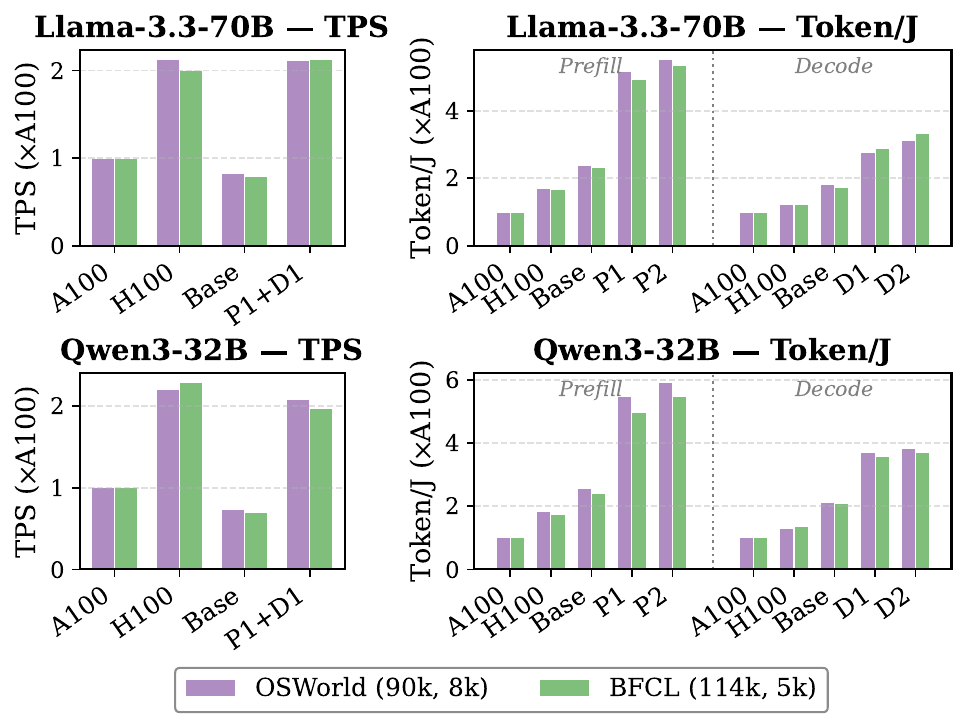}
    \caption{The P1 configuration achieves low TTFT, although it remains higher than A100 and H100 due to reduced compute resources. The D1 configuration delivers higher TPS, improving decoding throughput. The P2 and D2 configurations prioritize energy efficiency, achieving significantly higher tokens-per-joule. Finally, the combined P1 + D1 configuration provides a balanced design, achieving both low TTFT and high TPS across prefill and decode stages. For the PD disaggregation modelling, the communication channel is modelled with the NVLink, similar to the LLMCompass~\cite{llmcompass}.}
    \label{fig:sys_perf}
\end{figure}

To evaluate system performance, we maximize the batch size supported by the available memory hierarchy. We compare against A100 and H100 GPUs. Since a single device cannot accommodate large models under agentic workloads, we use four GPUs for each platform. For fairness, we compare four PLENA devices with the optimized memory system against four A100 and four H100 GPUs. Results are in \Cref{fig:sys_perf}.

Taken together, these results show that the benefit of PD disaggregation on heterogeneous NPUs comes not from the prefill/decode split itself, which prior systems~\cite{distserve, Splitwise} already exploit on homogeneous GPUs, but from \emph{which} technologies the co-design assigns to each stage. Prefill, dominated by large activation footprints, favors large on-chip SRAM (or 3D-stacked SRAM) with large tile sizes and an activation-stationary strategy that keeps activations on chip while streaming weights (P1, P2 in \Cref{tab:sample_config}), cutting off-chip traffic for reused data and relaxing the need for extreme HBM bandwidth. Decode, bounded by KV-cache capacity and off-chip bandwidth, favors high-capacity yet high-bandwidth HBF backed by energy-efficient LPDDR tiers and flatter systolic-array shapes (D1, D2), raising KV-cache residency and batch size while chunk-based double buffering overlaps data transfer with computation. Selecting these per-stage technologies, jointly with compute and data movement, delivers up to 6.8$\times$ higher TPS and 9.1$\times$ higher Token/J over Base in \Cref{fig:codesign_improvement}. Therefore, we draw the following key insight.

{\textbf{Key Insight 1}: Under PD disaggregation, prefill benefits from large on-chip SRAM to improve performance, whereas decode benefits from high-capacity, energy-efficient off-chip memory to improve throughput and power efficiency.}

\subsection{Working with emerging AI models}

\subsubsection{Diffusion Language Model (dLLMs)} dLLMs generate tokens through iterative denoising over the entire sequence, instead of autoregressive token-by-token decoding. During inference, the model repeatedly processes the full sequence across multiple diffusion steps, resulting in repeated full-sequence computation and memory accesses. Compared with autoregressive models, dLLMs exhibit larger activation footprints, since intermediate activations scale with sequence length and must be reused across multiple diffusion iterations. As a result, dLLMs place higher demands on on-chip memory capacity and bandwidth to efficiently store and reuse activations~\cite{llada}.

Since long-context agentic workloads are not well supported in current dLLM models~\cite{llada}, we adopt the GSM8K benchmark~\cite{gsm8k} (1.4K prompt, 0.2K generation) as a representative workload. The search results are shown in \Cref{tab:diffusion_case}.

\begin{table}[t]
\centering
\caption{Performance and energy efficiency comparison for LLaDA-8B (Diffusion)~\cite{llada} under different memory configurations, normalized to the baseline. As shown, storing all activations and a portion of the KV cache on chip leads to significant performance improvements. Hence, both prefill and decode tend to convert to a similar memory design.}
\label{tab:diffusion_case}
\setlength{\tabcolsep}{3pt}
\small
\begin{tabular}{lccc}
\toprule

& Baseline
& Prefill Optimized
& Decode Optimized \\
\midrule

On-chip      & SRAM   & 3D-Stack $\times$ 2 & 3D-Stack $\times$ 3  \\
L1 Off-chip  & HBM3E $\times$ 4 & HBM3E $\times$ 2  & HBM3E $\times$ 2 \\

\midrule
Power        & 281.89 & 307.76 & 341.21 \\
Batch Size   & 128 & 64 & 64 \\
Token/J      & 1.00$\times$ & \best{1.65$\times$} & \best{1.33$\times$} \\
\bottomrule
\end{tabular}
\end{table}

\subsubsection{Large MoE Model with Sparsity}

To investigate memory hierarchy design for extremely large sparse Mixture-of-Experts (MoE) models, we include Qwen3.5-397B-A17B~\cite{qwen3.5} as a representative case study. This model contains 397B total parameters, requiring approximately 370\,GB of storage for weights, while activating only 17B parameters per token. Such characteristics make it representative of large-scale sparse inference workloads, where only a subset of experts is activated at each step. Supporting this large-scale sparse model requires substantial off-chip memory capacity alongside high memory bandwidth to sustain expert loading and token routing.

According to the search results, HBF is frequently selected as part of the optimal configurations due to its high capacity and moderate-to-high bandwidth, making it well-suited for storing infrequently accessed expert weights while maintaining strong performance. Additionally, intermediate activations in this large-scale MoE model are substantial. By adopting 3D-stacked SRAM as on-chip memory, we significantly reduce off-chip traffic and improve data reuse (See \Cref{tab:large_moe_case}).

\begin{table}[t]
\centering
\caption{Performance and energy efficiency comparison for Qwen3.5-397B-A17B (MoE) under different memory configurations, normalized to the baseline PLENA + HBF$\times$2 configuration.}
\label{tab:large_moe_case}
\setlength{\tabcolsep}{3pt}
\small
\begin{tabular}{lccc}
\toprule

& Baseline
& Prefill Optimized
& Decode Optimized \\
\midrule

On-chip      & SRAM            & 3D-SRAM$\times$4 & SRAM \\
L1 Off-chip  & HBF$\times$2   & HBF$\times$2    & HBF$\times$1 \\
L2 Off-chip  & —               & —               & LPDDR5X$\times$8 \\
L3 Off-chip  & —               & —               & LPDDR5X$\times$8 \\

\midrule
Power        & 519.42          & 648.93          & \best{441.09} \\
Batch Size   & 32              & 128             & 64 \\
Token/J          & 1.00$\times$    & \best{3.52$\times$}    & 1.13$\times$ \\
\bottomrule
\end{tabular}
\end{table}

\subsection{Towards Extreme Heterogeneity}

The overall inference process can be decomposed into multiple stages. In this work, we investigate the potential of exploiting the extreme heterogeneity of the memory system to maximize performance, beyond conventional prefill and decode disaggregation. Concretely, we explore two finer-grained decompositions. First, we split each phase at the layer level into attention and FFN, so that the two can be mapped to different system configurations. Second, since the decode stage runs over many iterations whose memory traffic grows as the KV cache accumulates, we partition decode into multiple temporal phases based on the progression of generated tokens.

\begin{figure}[t]
    \centering
    \includegraphics[width=\linewidth]{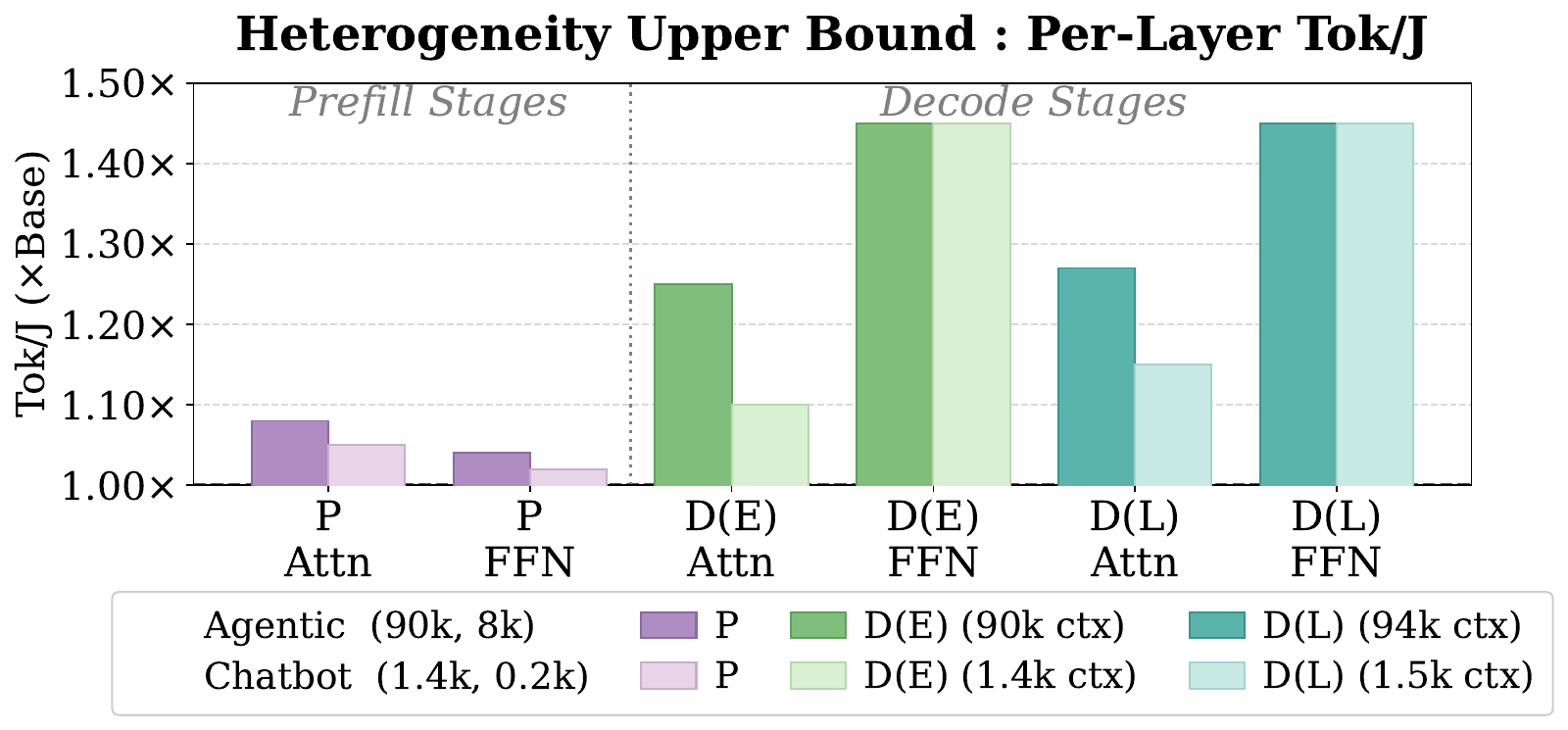}
    \caption{ Per-layer performance comparison of the NPU designs optimized separately for attention and FFN across three phases: prefill, early decode (first 50\% of generated tokens), and late decode (remaining 50\%), using the P1 and D1 configurations from \Cref{tab:sample_config} for prefill and decode performance, respectively. Along the x-axis, \textbf{P} denotes the prefill phase, \textbf{D(E)} the early-decode phase, and \textbf{D(L)} the late-decode phase, each split into its attention (Attn) and FFN sub-layers; results are reported for two workloads, an agentic ($90$k input, $8$k output) and a chatbot ($1.4$k input, $0.2$k output) trace. Values are Tok/J normalized to the homogeneous baseline ($\times$Base: P1 for prefill, D1 for decode), so a value above $1.0\times$ indicates a gain from per-layer heterogeneity.}
    \label{fig:heterogenuity_upperbound}
\end{figure}

According to \Cref{fig:heterogenuity_upperbound}, the benefits of attention--FFN specialization vary substantially across serving phases. During prefill, both attention and FFN are dominated by large, highly parallel matrix operations and therefore favor similar compute-oriented configurations. Mapping them to separate designs consequently provides little benefit. During decode, however, the two diverge sharply: attention is bound by repeated KV-cache accesses and favors high-capacity, bandwidth-efficient memory (e.g., HBF or LPDDR), whereas FFN remains compute-intensive and prefers compute-dense engines with weight streaming. Mapping the two to separate engines therefore improves decode efficiency, consistent with disaggregated designs such as NVIDIA Vera Rubin paired with Groq LPX~\cite{nvidia_vera_rubin_platform}.
By contrast, optimizing early and late decode separately provides no consistent improvement across the evaluated workloads. Although KV-cache traffic grows during generation, this change is insufficient to alter the preferred system configuration within a single request.

{\textbf{Key Insight 2}: Attention--FFN disaggregation offers limited benefit during prefill but improves decode efficiency by specializing attention for KV-cache-intensive execution and FFN for compute-intensive execution. Further temporal decomposition of decode provides little additional benefit for the evaluated workloads in the prefill-heavy agentic task.}

\subsection{Model Validation}

We validate MemExplorer along three axes, latency, power, and area, each against independent ground truth: a cycle-accurate emulator, RTL synthesis, and real GPU devices.

{\textbf{Performance Validation}}
For latency, our reference is the PLENA~\cite{plena} transactional emulator,
which models off-chip memory behavior with Ramulator~\cite{ramulator} and has
itself been validated on FPGA platforms with HBM. We extend it to cover the
technologies MemExplorer targets, namely 3D-stacked SRAM and HBF, and then run
both tools under identical hardware configurations and workloads.

We use only the GSM8K trace because the cycle-accurate PLENA emulator incurs substantial simulation time, requiring several hours for each decode sequence. We therefore select the relatively short GSM8K workload~\cite{gsm8k} for validation. The results show that MemExplorer closely matches the emulator, with an overall error below 14\%.

However, agreement with an emulator alone does not establish that the model accurately reflects real hardware. We therefore also validate MemExplorer against commercial GPUs. Each GPU is represented using a roofline model integrated into MemExplorer, and the predicted prefill latency, decode latency, and end-to-end throughput are compared with measurements across four devices spanning HBM2e, HBM3, HBM3e, (\Cref{tab:gpu_validation}).

\begin{figure}[t]
    \centering
    \includegraphics[width=\linewidth]{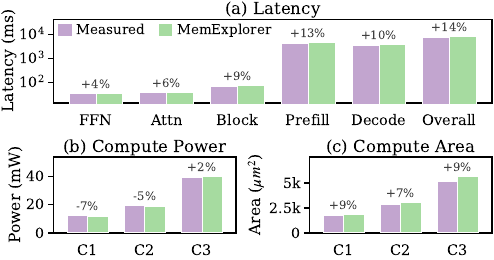}
    \caption{Validation of MemExplorer against ground truth: measured versus predicted values, with the signed relative error above each pair. \emph{(a)}~Latency against the cycle-accurate PLENA emulator for Qwen3-32B (GSM8K traces of 1.4K/0.2K under the P1 configuration), log scale; FFN/Attn are the feed-forward and attention sub-layers, Block is one complete Transformer layer, and Prefill/Decode/Overall are the full-process stages. \emph{(b,c)}~Compute PE-array power and area against RTL synthesis results from Synopsys DC, for configurations C1--C3 $=4{\times}16$/$\mathrm{VLEN}16$, $4{\times}32$/$\mathrm{VLEN}32$, $64{\times}64$/$\mathrm{VLEN}64$ (PE-array size / vector-lane count), each including the HBM interface and on-chip SRAM. Across the emulator, power, and area points the mean absolute error is 7.9\% (max 14.5\%).}
    \label{fig:validation_parity}
\end{figure}

\textbf{Power and area validation.}
They are validated separately for compute and for memory. On the
compute side, we synthesize the PE array with Synopsys Design Compiler using the
7\,nm OpenROAD predictive PDK~\cite{openroad}, sweeping three representative
configurations, \emph{C1}--\emph{C3}. \Cref{fig:validation_parity}(b,c) report the
resulting power and area. On the memory side, we compare predicted HBM device
power against sequential-access measurements on an AMD Alveo V80 FPGA under
write-only, read-only, and alternating read/write patterns
(\Cref{tab:hbm_power_validation}).

Overall, validation against the emulator, Synopsys synthesis results, real GPUs, and measured HBM power establishes the accuracy and reliability of the MemExplorer model.

\begin{table}[t]
  \centering
  \caption{GPU validation across different memory technologies on OSWorld with
  90K input tokens and 8K output tokens. We report measured and simulated
  prefill latency, decode latency, and end-to-end TPS. All experiments use
  Qwen3-32B in FP8 with batch size 2, except A100, which uses BF16 with batch
  size 1.}
  \label{tab:gpu_validation}
  \small
  \renewcommand{\arraystretch}{1.1}
  \setlength{\tabcolsep}{1.3pt}

  \begin{tabular}{lcccccc}
  \toprule
  \multirow{2}{*}{\textbf{GPU}} &
  \multicolumn{2}{c}{\textbf{Prefill Latency}} &
  \multicolumn{2}{c}{\textbf{Decode Latency}} &
  \multicolumn{2}{c}{\textbf{End-to-End TPS}} \\
  \cmidrule(lr){2-3}
  \cmidrule(lr){4-5}
  \cmidrule(lr){6-7}
  &
  \textbf{Real} &
  \textbf{Sim.} &
  \textbf{Real} &
  \textbf{Sim.} &
  \textbf{Real} &
  \textbf{Sim.} \\
  \midrule

  A100$^\ddagger$
  & 92.5
  & 84.0 {\scriptsize($-$9.2\%)}
  & 362.2
  & 380.7 {\scriptsize($+$5.1\%)}
  & 17.6
  & 17.2 {\scriptsize($-$2.2\%)} \\

  6000$^\dagger$
  & 186.8
  & 177.8 {\scriptsize($-$4.8\%)}
  & 501.0
  & 479.3 {\scriptsize($-$4.3\%)}
  & 23.3
  & 24.4 {\scriptsize($+$4.7\%)} \\

  H100
  & 31.3
  & 32.0 {\scriptsize($+$2.5\%)}
  & 280.9
  & 245.3 {\scriptsize($-$12.7\%)}
  & 51.3
  & 57.7 {\scriptsize($+$12.6\%)} \\

  B200
  & 23.2
  & 20.1 {\scriptsize($-$13.3\%)}
  & 178.8
  & 176.1 {\scriptsize($-$1.5\%)}
  & 79.2
  & 81.6 {\scriptsize($+$3.0\%)} \\

  \bottomrule
  \end{tabular}

  \vspace{0.2em}
  \begin{flushleft}
  \footnotesize
  $^\dagger$6000 denotes RTX Pro 6000.
  A100: HBM2e 80GB; H100: HBM3 80GB; RTX Pro 6000: GDDR7 96GB;
  B200: HBM3e 180GB.
  $^\ddagger$A100 uses BF16 because it does not support FP8 computation.
  \end{flushleft}
\end{table}

\begin{table}[t]
  \centering
  \caption{HBM power validation against sequential-access measurements on an AMD Alveo V80 FPGA, under write-only, read-only, and alternating
  read/write access.}
  \label{tab:hbm_power_validation}
  \small
  \renewcommand{\arraystretch}{1.1}
  \setlength{\tabcolsep}{4pt}
  \begin{tabular}{lccc}
  \toprule
  \textbf{Access Pattern} &
  \textbf{Measured (W)} &
  \textbf{MemExplorer (W)} &
  \textbf{Error} \\
  \midrule
  Write-only  & 103.37 & 115.70 & $+$11.9\% \\
  Read-only   & 108.16 & 112.30 & $+$3.8\%  \\
  Read/Write  & 106.99 & 114.00 & $+$6.6\%  \\
  \bottomrule
  \end{tabular}
\end{table}

\section{Related Work}

{\bf Memory technologies.}
As shown in \Cref{fig:mem_design_space}, a wide range of memory technologies has emerged to meet the growing capacity and bandwidth demands of modern AI accelerators~\cite{YuK24}, yet none simultaneously provides high bandwidth, large capacity, and low power. HBM~\cite{hbm_intro} is widely used in high-performance accelerators from NVIDIA V100~\cite{v100} to B200~\cite{b200} and upcoming Vera Rubin systems~\cite{nvidia_vera_rubin_platform}, while GDDR~\cite{gddr_intro,moon_jedec_gddr7_2025} provides a cost-effective alternative. LPDDR~\cite{micron_lpddr5x,synopsys_lpddr6_vs_lpddr5x_2026} targets energy-efficient deployments, including datacenter platforms such as NVIDIA Grace~\cite{lpddr5x_power}.

Recently, HBF~\cite{kim_hbf_roadmap_2026,HBF} has emerged to bridge the capacity gap between high-bandwidth memory and traditional storage.
On the on-chip side, conventional 2D SRAM provides the highest bandwidth with sub-nanosecond latency but is limited to hundreds of megabytes per die due to area and leakage constraints. 3D-stacked SRAM, as demonstrated by AMD's V-Cache~\cite{wuu20223d}, circumvents this by bonding multiple SRAM dies vertically, significantly expanding on-chip capacity without increasing die footprint. These diverse trade-offs across bandwidth, capacity, power, and physical constraints necessitate heterogeneous memory hierarchies, whose detailed characteristics we model in \Cref{tab:offchip_memory_params}.

{\bf Hardware--Software Co-design on Memory Technologies.}
Accurate modeling of such hierarchies remains an open challenge.
Existing memory simulators, including Ramulator~\cite{ramulator} and DRAMSys~\cite{dramsys}, provide detailed cycle-accurate simulation of DRAM behavior, capturing timing, scheduling, and power characteristics at the device level.
However, these tools are designed for conventional DRAM technologies and do not model emerging memory technologies such as 3D-stacked SRAM or HBF. They do not capture end-to-end LLM inference workloads or accelerator-level execution characteristics, making them insufficient for evaluating memory-centric design trade-offs across heterogeneous hierarchies.

H3~\cite{HBF} proposes a hybrid HBM--HBF architecture for cost-efficient LLM inference and demonstrates the benefit of tiered off-chip memory, yet does not model on-chip memory hierarchies or detailed memory traffic patterns. Li et al.~\cite{li2026hardwaresoftwarecodesign3ddrambasedllm} present a hardware--software co-design for 3D-DRAM-based LLM serving accelerators, focusing on a specific 3D-DRAM technology rather than providing a general framework that spans diverse memory technologies.
MemExplorer bridges this gap by jointly modeling heterogeneous compute and memory hierarchies---spanning various technologies including the emerging 3D-stacked SRAM and HBF---enabling systematic exploration for modern LLM inference workloads.

{\bf Disaggregated serving of heterogeneous NPUs.}
To optimize LLM serving, PD disaggregation separates compute-intensive prefill from memory-bound decoding to reduce resource interference, as demonstrated by DistServe~\cite{distserve}, Splitwise~\cite{Splitwise}, Mooncake~\cite{mooncake}.
Recent NPU designs, such as RPU~\cite{rpu} and PLENA~\cite{plena}, explore disaggregation across heterogeneous NPU configurations. However, these works primarily focus on compute heterogeneity while assuming fixed memory configurations, limiting their ability to capture the memory requirements of prefill and decode stages.

\section{Conclusion}

The increasing hardware accelerator heterogeneity in emerging LLM systems leads to challenges in memory system architecture co-design.
We present MemExplorer, a novel memory system synthesizer that automatically identifies an efficient memory architecture for heterogeneous NPU systems.
We propose a unified abstraction for the systematic exploration of accelerator memory architecture and develop automated techniques that combine the exploration of both compute and memory design choices.

As the first attempt at memory architecture exploration for heterogeneous accelerator systems, MemExplorer has two main limitations. First, it does not model multi-device systems with shared memory, as chip-to-chip (C2C) interconnects and cross-device communication overhead are not currently captured. Second, it does not yet model memory write endurance or thermal effects, which become important for emerging memory technologies like 3D Stacked SRAM and HBF. Third, the precision is consistent and does not support mixed precision across the prefill and decode stages.
These will be our key future work to explore both theoretically and practically.
We plan to validate our framework on real GPU and accelerator platforms, extend the framework to multi-core compute systems with shared memory hierarchies, and incorporate mixed-precision execution and emerging memory technologies into the exploration.

\bibliographystyle{IEEEtranS}
\bibliography{sample-base}

@article{emmerich2006single,
  title={Single-and multiobjective evolutionary optimization assisted by Gaussian random field metamodels},
  author={Emmerich, Michael TM and Giannakoglou, Kyriakos C and Naujoks, Boris},
  journal={IEEE Transactions on Evolutionary Computation (TEC)},
  volume={10},
  number={4},
  pages={421--439},
  year={2006},
  publisher={IEEE}
}

@article{daulton2020differentiable,
  title={Differentiable expected hypervolume improvement for parallel multi-objective Bayesian optimization},
  author={Daulton, Samuel and Balandat, Maximilian and Bakshy, Eytan},
  journal={Advances in neural information processing systems (NeurIPS)},
  volume={33},
  pages={9851--9864},
  year={2020}
}

@inproceedings{
bfcl,
title={The Berkeley Function Calling Leaderboard ({BFCL}): From Tool Use to Agentic Evaluation of Large Language Models},
author={Shishir G Patil and Huanzhi Mao and Fanjia Yan and Charlie Cheng-Jie Ji and Vishnu Suresh and Ion Stoica and Joseph E. Gonzalez},
booktitle={Forty-second International Conference on Machine Learning},
year={2025},
url={https://openreview.net/forum?id=2GmDdhBdDk}
}

@misc{osworld,
      title={OSWorld: Benchmarking Multimodal Agents for Open-Ended Tasks in Real Computer Environments}, 
      author={Tianbao Xie and Danyang Zhang and Jixuan Chen and Xiaochuan Li and Siheng Zhao and Ruisheng Cao and Toh Jing Hua and Zhoujun Cheng and Dongchan Shin and Fangyu Lei and Yitao Liu and Yiheng Xu and Shuyan Zhou and Silvio Savarese and Caiming Xiong and Victor Zhong and Tao Yu},
      year={2024},
      eprint={2404.07972},
      archivePrefix={arXiv},
      primaryClass={cs.AI},
      url={https://arxiv.org/abs/2404.07972}, 
}

@ARTICLE{HBF,
  author={Ha, Minho and Kim, Euiseok and Kim, Hoshik},
  journal={IEEE Computer Architecture Letters}, 
  title={H3: Hybrid Architecture Using High Bandwidth Memory and High Bandwidth Flash for Cost-Efficient LLM Inference}, 
  year={2026},
  volume={25},
  number={1},
  pages={49-52},
  doi={10.1109/LCA.2026.3660969}}

@misc{micron_gddr6_flyer,
  author       = {{Micron Technology, Inc.}},
  title        = {{Micron GDDR6 Memory Product Flyer}},
  year         = {2019},
  howpublished = {\url{https://www.mouser.co.uk/pdfDocs/gddr6_product_flyer.pdf}},
  note         = {Technical product flyer, accessed: 2026-03-14}
}

@misc{micron_lpddr5x,
  author       = {{Micron Technology, Inc.}},
  title        = {{LPDDR5X DRAM Components}},
  year         = {2025},
  howpublished = {\url{https://www.micron.com/products/memory/dram-components/lpddr5x/}},
  note         = {Micron product page, accessed: 2026-03-14}
}

@misc{kim_hbf_roadmap_2026,
  author       = {Joungho Kim},
  title        = {HBF Technology, Workload Analysis and Roadmap},
  year         = {2026},
  howpublished = {Presentation slides, TeraByte Interconnection and Package Laboratory (TERALAB), KAIST},
  url          = {https://www.dropbox.com/scl/fo/pv1wqpyvqjz7qdra1z94e/AFeIVtavoj2HDWxbLzzYN7o/HBF_Workload_and_Roadmap_Joungho_Kim.pdf},
  note         = {School of Electrical Engineering, KAIST. Accessed: 2026-03-14}
}

@misc{synopsys_lpddr6_vs_lpddr5x_2026,
  author       = {Scott Knowlton},
  title        = {LPDDR6 vs. LPDDR5 and LPDDR5X: What’s the Difference?},
  year         = {2026},
  month        = feb,
  howpublished = {Synopsys Blog},
  url          = {https://www.synopsys.com/blogs/chip-design/lpddr6-vs-lpddr5x-lpddr5-differences.html},
  note         = {Accessed: 2026-03-14}
}

@misc{lpddr5x_power,
  author       = {Harris, Dion},
  title        = {NVIDIA Grace and Grace Hopper Update},
  howpublished = {Presentation at the HPC User Forum, Argonne National Laboratory},
  year         = {2023},
  month        = {September},
  day          = {7},
  url          = {https://www.hpcuserforum.com/wp-content/uploads/2023/09/Dion-Harris-Update-on-NVIDIA-Grace_Grace-Hopper.2.pdf},
  note         = {Slide 15: "Energy Efficient Design" highlights LPDDR5X at 5 pJ/bit vs. DDR5 at 35 pJ/bit.}
}

@misc{techpowerup_gddr7_voltage_2023,
  title        = {Samsung GDDR7 Memory Operates at Lower Voltage, Built on Same Node as 24 Gbps GDDR6},
  author       = {{TechPowerUp}},
  year         = {2023},
  howpublished = {\url{https://www.techpowerup.com/311457/samsung-gddr7-memory-operates-at-lower-voltage-built-on-same-node-as-24-gbps-g6}},
  note         = {Accessed: 2026-03-16}
}

@techreport{moon_jedec_gddr7_2025,
  author      = {Seunghyun Moon},
  title       = {Next-Generation GDDR7 Memory Technology},
  institution = {JEDEC},
  year        = {2025},
  month       = mar,
  url         = {https://www.jedec.org/sites/default/files/Seunghyun%20Moon_03_29_25_Final.pdf},
  note        = {JEDEC Presentation}
}

@online{emergentmind_hbf_2026,
  author  = {{Emergent Mind}},
  title   = {High Bandwidth Flash (HBF) Overview},
  year    = {2026},
  url     = {https://www.emergentmind.com/topics/high-bandwidth-flash-hbf},
  note    = {Updated Jan. 12, 2026. Accessed: 2026-03-16}
}

@misc{patsnap_hbm_wars_2025,
  author       = {{PatSnap Eureka}},
  title        = {The HBM Wars: SK Hynix's Dominance, Samsung's Roadmap, and the Looming Threat of Cyclicality},
  year         = {2025},
  month        = aug,
  howpublished = {\url{https://eureka.patsnap.com/insight/the-hbm-wars-sk-hynixs-dominance-samsungs-roadmap-and-the-looming-threat-of-cyclicality}},
  note         = {Accessed: 2026-03-16}
}

@article{ma2026llmhardware,
  title   = {Challenges and Research Directions for Large Language Model Inference Hardware},
  author  = {Ma, Xiaoyu and Patterson, David},
  journal = {arXiv preprint arXiv:2601.05047},
  year    = {2026},
  doi     = {10.48550/arXiv.2601.05047}
}

@misc{longbench,
      title={LongBench v2: Towards Deeper Understanding and Reasoning on Realistic Long-context Multitasks}, 
      author={Yushi Bai and Shangqing Tu and Jiajie Zhang and Hao Peng and Xiaozhi Wang and Xin Lv and Shulin Cao and Jiazheng Xu and Lei Hou and Yuxiao Dong and Jie Tang and Juanzi Li},
      year={2025},
      eprint={2412.15204},
      archivePrefix={arXiv},
      primaryClass={cs.CL},
      url={https://arxiv.org/abs/2412.15204}, 
}

@misc{plena,
      title={Combating the Memory Walls: Optimization Pathways for Long-Context Agentic LLM Inference}, 
      author={Haoran Wu and Can Xiao and Jiayi Nie and Xuan Guo and Binglei Lou and Jeffrey T. H. Wong and Zhiwen Mo and Cheng Zhang and Przemyslaw Forys and Wayne Luk and Hongxiang Fan and Jianyi Cheng and Timothy M. Jones and Rika Antonova and Robert Mullins and Aaron Zhao},
      year={2025},
      eprint={2509.09505},
      archivePrefix={arXiv},
      primaryClass={cs.AR},
      url={https://arxiv.org/abs/2509.09505}, 
}

@misc{v100,
  title        = {{NVIDIA Tesla V100 GPU Architecture Datasheet}},
  author       = {{NVIDIA Corporation}},
  year         = {2018},
  howpublished = {\url{https://images.nvidia.com/content/technologies/volta/pdf/volta-v100-datasheet-update-us-1165301-r5.pdf}},
  note         = {Version R5}
}

@misc{b200,
  title        = {{NVIDIA B200 GPU Specifications}},
  author       = {{TechPowerUp}},
  year         = {2025},
  howpublished = {\url{https://www.techpowerup.com/gpu-specs/b200.c4210}},
  note         = {Accessed: 2026-03-22}
}

@misc{ouyang2025kernelbenchllmswriteefficient,
      title={KernelBench: Can LLMs Write Efficient GPU Kernels?}, 
      author={Anne Ouyang and Simon Guo and Simran Arora and Alex L. Zhang and William Hu and Christopher Ré and Azalia Mirhoseini},
      year={2025},
      eprint={2502.10517},
      archivePrefix={arXiv},
      primaryClass={cs.LG},
      url={https://arxiv.org/abs/2502.10517}, 
}

@misc{kernelcraft,
      title={KernelCraft: Benchmarking for Agentic Close-to-Metal Kernel Generation on Emerging Hardware}, 
      author={Jiayi Nie and Haoran Wu and Yao Lai and Zeyu Cao and Cheng Zhang and Binglei Lou and Erwei Wang and Jianyi Cheng and Timothy M. Jones and Robert Mullins and Rika Antonova and Yiren Zhao},
      year={2026},
      eprint={2603.08721},
      archivePrefix={arXiv},
      primaryClass={cs.AR},
      url={https://arxiv.org/abs/2603.08721}, 
}

@misc{agent_s2,
      title={Agent S2: A Compositional Generalist-Specialist Framework for Computer Use Agents}, 
      author={Saaket Agashe and Kyle Wong and Vincent Tu and Jiachen Yang and Ang Li and Xin Eric Wang},
      year={2025},
      eprint={2504.00906},
      archivePrefix={arXiv},
      primaryClass={cs.AI},
      url={https://arxiv.org/abs/2504.00906}, 
}

@misc{webagent,
      title={Web Agents with World Models: Learning and Leveraging Environment Dynamics in Web Navigation}, 
      author={Hyungjoo Chae and Namyoung Kim and Kai Tzu-iunn Ong and Minju Gwak and Gwanwoo Song and Jihoon Kim and Sunghwan Kim and Dongha Lee and Jinyoung Yeo},
      year={2025},
      eprint={2410.13232},
      archivePrefix={arXiv},
      primaryClass={cs.CL},
      url={https://arxiv.org/abs/2410.13232}, 
}

@misc{llmcompass,
      title={A Hardware Evaluation Framework for Large Language Model Inference}, 
      author={Hengrui Zhang and August Ning and Rohan Prabhakar and David Wentzlaff},
      year={2023},
      eprint={2312.03134},
      archivePrefix={arXiv},
      primaryClass={cs.AR},
      url={https://arxiv.org/abs/2312.03134}, 
}

@misc{rpu,
      title={RPU -- A Reasoning Processing Unit}, 
      author={Matthew Adiletta and Gu-Yeon Wei and David Brooks},
      year={2026},
      eprint={2602.18568},
      archivePrefix={arXiv},
      primaryClass={cs.AR},
      url={https://arxiv.org/abs/2602.18568}, 
}

@INPROCEEDINGS{hbm_intro,
  author={Jun, Hongshin and Cho, Jinhee and Lee, Kangseol and Son, Ho-Young and Kim, Kwiwook and Jin, Hanho and Kim, Keith},
  booktitle={2017 IEEE International Memory Workshop (IMW)}, 
  title={HBM (High Bandwidth Memory) DRAM Technology and Architecture}, 
  year={2017},
  volume={},
  number={},
  pages={1-4},
  doi={10.1109/IMW.2017.7939084}}

@misc{gddr_intro,
  title        = {GDDR6 SGRAM Standard},
  author       = {{JEDEC Solid State Technology Association}},
  year         = {2023},
  note         = {JESD250},
  url          = {https://www.jedec.org/standards-documents/docs/jesd250}
}

@misc{nvidia_vera_rubin_platform,
  title        = {NVIDIA Vera Rubin Platform Opens the Next Frontier of Agentic AI},
  author       = {{NVIDIA Corporation}},
  year         = {2026},
  howpublished = {\url{https://nvidianews.nvidia.com/news/nvidia-vera-rubin-platform}},
  note         = {Press release, Accessed: 2026-03-22}
}

@article{llama3,
  title={Introducing meta llama 3: The most capable openly available llm to date},
  author={Meta, AI},
  journal={Meta AI},
  year={2024}
}

@misc{qwen3,
      title={Qwen3 Technical Report}, 
      author={An Yang and Anfeng Li and Baosong Yang and Beichen Zhang and Binyuan Hui and Bo Zheng and Bowen Yu and Chang Gao and Chengen Huang and Chenxu Lv and Chujie Zheng and Dayiheng Liu and Fan Zhou and Fei Huang and Feng Hu and Hao Ge and Haoran Wei and Huan Lin and Jialong Tang and Jian Yang and Jianhong Tu and Jianwei Zhang and Jianxin Yang and Jiaxi Yang and Jing Zhou and Jingren Zhou and Junyang Lin and Kai Dang and Keqin Bao and Kexin Yang and Le Yu and Lianghao Deng and Mei Li and Mingfeng Xue and Mingze Li and Pei Zhang and Peng Wang and Qin Zhu and Rui Men and Ruize Gao and Shixuan Liu and Shuang Luo and Tianhao Li and Tianyi Tang and Wenbiao Yin and Xingzhang Ren and Xinyu Wang and Xinyu Zhang and Xuancheng Ren and Yang Fan and Yang Su and Yichang Zhang and Yinger Zhang and Yu Wan and Yuqiong Liu and Zekun Wang and Zeyu Cui and Zhenru Zhang and Zhipeng Zhou and Zihan Qiu},
      year={2025},
      eprint={2505.09388},
      archivePrefix={arXiv},
      primaryClass={cs.CL},
      url={https://arxiv.org/abs/2505.09388}, 
}

@article{openroad,
  author    = {L. T. Clark and V. Vashishtha and L. Shifren and A. Gujja and S. Sinha and B. Cline and C. Ramamurthy and G. Yeric},
  title     = {ASAP: A 7-nm finFET predictive process design kit},
  journal   = {Microelectronics Journal},
  volume    = {53},
  pages     = {105--115},
  year      = {2016},
  month     = jul,
  doi       = {10.1016/j.mejo.2016.04.006}
}

@misc{muchisim,
      title={Muchisim: A Simulation Framework for Design Exploration of Multi-Chip Manycore Systems}, 
      author={Marcelo Orenes-Vera and Esin Tureci and Margaret Martonosi and David Wentzlaff},
      year={2024},
      eprint={2312.10244},
      archivePrefix={arXiv},
      primaryClass={cs.AR},
      url={https://arxiv.org/abs/2312.10244}, 
}

@misc{waferscale,
      title={Theseus: Exploring Efficient Wafer-Scale Chip Design for Large Language Models}, 
      author={Jingchen Zhu and Chenhao Xue and Yiqi Chen and Zhao Wang and Chen Zhang and Yu Shen and Yifan Chen and Zekang Cheng and Yu Jiang and Tianqi Wang and Yibo Lin and Wei Hu and Bin Cui and Runsheng Wang and Yun Liang and Guangyu Sun},
      year={2024},
      eprint={2407.02079},
      archivePrefix={arXiv},
      primaryClass={cs.AR},
      url={https://arxiv.org/abs/2407.02079}, 
}

@misc{asml,
  title = {ASML TWINSCAN NXE EUV Lithography Systems},
  author = {ASML},
  year = {2024},
  url = {https://www.asml.com/products/euv-lithography-systems/twinscan-nxe3400b},
  note = {Maximum exposure field size of 26 mm × 33 mm}
}

@article{kimi1m,
  title={Kimi Linear: An Expressive, Efficient Attention Architecture},
  author={{Moonshot AI}},
  journal={arXiv preprint arXiv:2412.18525},
  year={2024}
}

@misc{glm4,
      title={ChatGLM: A Family of Large Language Models from GLM-130B to GLM-4 All Tools}, 
      author={Team GLM and : and Aohan Zeng and Bin Xu and Bowen Wang and Chenhui Zhang and Da Yin and Dan Zhang and Diego Rojas and Guanyu Feng and Hanlin Zhao and Hanyu Lai and Hao Yu and Hongning Wang and Jiadai Sun and Jiajie Zhang and Jiale Cheng and Jiayi Gui and Jie Tang and Jing Zhang and Jingyu Sun and Juanzi Li and Lei Zhao and Lindong Wu and Lucen Zhong and Mingdao Liu and Minlie Huang and Peng Zhang and Qinkai Zheng and Rui Lu and Shuaiqi Duan and Shudan Zhang and Shulin Cao and Shuxun Yang and Weng Lam Tam and Wenyi Zhao and Xiao Liu and Xiao Xia and Xiaohan Zhang and Xiaotao Gu and Xin Lv and Xinghan Liu and Xinyi Liu and Xinyue Yang and Xixuan Song and Xunkai Zhang and Yifan An and Yifan Xu and Yilin Niu and Yuantao Yang and Yueyan Li and Yushi Bai and Yuxiao Dong and Zehan Qi and Zhaoyu Wang and Zhen Yang and Zhengxiao Du and Zhenyu Hou and Zihan Wang},
      year={2024},
      eprint={2406.12793},
      archivePrefix={arXiv},
      primaryClass={cs.CL},
      url={https://arxiv.org/abs/2406.12793}, 
}

@article{qwen1m,
  title={Qwen2.5-1M Technical Report},
  author={{Qwen Team}},
  journal={arXiv preprint arXiv:2501.15383},
  year={2025}
}

@misc{qwen3.5,
    title  = {{Qwen3.5}: Towards Native Multimodal Agents},
    author = {{Qwen Team}},
    month  = {February},
    year   = {2026},
    url    = {https://qwen.ai/blog?id=qwen3.5}
}

@article{YuK24,
  author          = {Shimeng Yu and Tae-Hyeon Kim},
  title           = {Semiconductor Memory Technologies: State-of-the-Art and Future Trends},
  journal         = {Computer},
  year            = {2024},
  volume          = {57},
  number          = {1},
  pages           = {150--154},
  doi             = {10.1109/MC.2023.3332791}
}

@INPROCEEDINGS{wuu20223d,
  author={Wuu, John and Agarwal, Rahul and Ciraula, Michael and Dietz, Carl and Johnson, Brett and Johnson, Dave and Schreiber, Russell and Swaminathan, Raja and Walker, Will and Naffziger, Samuel},
  booktitle={2022 IEEE International Solid-State Circuits Conference (ISSCC)}, 
  title={3D V-Cache: the Implementation of a Hybrid-Bonded 64MB Stacked Cache for a 7nm x86-64 CPU}, 
  year={2022},
  volume={65},
  number={},
  pages={428-429},
  doi={10.1109/ISSCC42614.2022.9731565}}

@misc{llada,
      title={Large Language Diffusion Models}, 
      author={Shen Nie and Fengqi Zhu and Zebin You and Xiaolu Zhang and Jingyang Ou and Jun Hu and Jun Zhou and Yankai Lin and Ji-Rong Wen and Chongxuan Li},
      year={2025},
      eprint={2502.09992},
      archivePrefix={arXiv},
      primaryClass={cs.CL},
      url={https://arxiv.org/abs/2502.09992}, 
}

@article{ramulator,
author = {Luo, Haocong and Tu\u{g}rul, Yahya Can and Bostanc\i{}, F. Nisa and Olgun, Ataberk and Ya\u{g}l\i{}k\c{c}\i{}, A. Giray and Mutlu, Onur},
title = {Ramulator 2.0: A Modern, Modular, and Extensible DRAM Simulator},
year = {2024},
issue_date = {Jan.-June 2024},
publisher = {IEEE Computer Society},
address = {USA},
volume = {23},
number = {1},
issn = {1556-6056},
url = {https://doi.org/10.1109/LCA.2023.3333759},
doi = {10.1109/LCA.2023.3333759},
journal = {IEEE Comput. Archit. Lett.},
month = jan,
pages = {112–116},
numpages = {5}
}

@article{dramsys,
author = {Steiner, Lukas and Jung, Matthias and Prado, Felipe S. and Bykov, Kirill and Wehn, Norbert},
title = {DRAMSys4.0: An Open-Source Simulation Framework for In-depth DRAM Analyses},
year = {2022},
issue_date = {Apr 2022},
publisher = {Kluwer Academic Publishers},
address = {USA},
volume = {50},
number = {2},
issn = {0885-7458},
url = {https://doi.org/10.1007/s10766-022-00727-4},
doi = {10.1007/s10766-022-00727-4},
journal = {Int. J. Parallel Program.},
month = apr,
pages = {217–242},
numpages = {26}
}

@misc{li2026hardwaresoftwarecodesign3ddrambasedllm,
      title={Hardware-Software Co-design for 3D-DRAM-based LLM Serving Accelerator}, 
      author={Cong Li and Yihan Yin and Chenhao Xue and Zhao Wang and Fujun Bai and Yixin Guo and Xiping Jiang and Qiang Wu and Yuan Xie and Guangyu Sun},
      year={2026},
      eprint={2603.04797},
      archivePrefix={arXiv},
      primaryClass={cs.AR},
      url={https://arxiv.org/abs/2603.04797}, 
}

@inproceedings{
gsm8k,
title={{MR}-{GSM}8K: A Meta-Reasoning Benchmark for Large Language Model Evaluation},
author={Zhongshen Zeng and Pengguang Chen and Shu Liu and Haiyun Jiang and Jiaya Jia},
booktitle={The Thirteenth International Conference on Learning Representations},
year={2025},
url={https://openreview.net/forum?id=br4H61LOoI}
}

@INPROCEEDINGS{dataflow,
  author={Lee, Jeong-Jun and Li, Peng},
  booktitle={2020 IEEE 38th International Conference on Computer Design (ICCD)}, 
  title={Reconfigurable Dataflow Optimization for Spatiotemporal Spiking Neural Computation on Systolic Array Accelerators}, 
  year={2020},
  volume={},
  number={},
  pages={57-64},
  doi={10.1109/ICCD50377.2020.00027}}

@INPROCEEDINGS{shorline,
  author={Chen, Victor and Abdel-Dayem, Bassem and Wan, Changhua and Ling, Feng},
  booktitle={2022 IEEE International Symposium on Electromagnetic Compatibility \& Signal/Power Integrity (EMCSI)}, 
  title={Overcoming Design Challenges for High Bandwidth Memory Interface with CoWoS}, 
  year={2022},
  volume={},
  number={},
  pages={455-458},
  doi={10.1109/EMCSI39492.2022.10050234}}

@misc{distserve,
      title={DistServe: Disaggregating Prefill and Decoding for Goodput-optimized Large Language Model Serving}, 
      author={Yinmin Zhong and Shengyu Liu and Junda Chen and Jianbo Hu and Yibo Zhu and Xuanzhe Liu and Xin Jin and Hao Zhang},
      year={2024},
      eprint={2401.09670},
      archivePrefix={arXiv},
      primaryClass={cs.DC},
      url={https://arxiv.org/abs/2401.09670}, 
}

@misc{cerebras_aws,
      title={{AWS} and Cerebras Collaboration Sets a New Standard for {AI} Inference Speed and Performance in the Cloud}, 
      author={Wang, James},
      year={2026},
      month={March},
      howpublished={\url{https://www.cerebras.ai/blog/cerebras-is-coming-to-aws}},
      note={Accessed: 2026-04-07}
}

@book{yu2022semiconductor,
  title={Semiconductor Memory Devices and Circuits},
  author={Yu, S.},
  isbn={9781000567618},
  lccn={2021053446},
  url={https://books.google.co.uk/books?id=FPBjEAAAQBAJ},
  year={2022},
  publisher={CRC Press}
}

@misc{Splitwise,
      title={Splitwise: Efficient generative LLM inference using phase splitting}, 
      author={Pratyush Patel and Esha Choukse and Chaojie Zhang and Aashaka Shah and Íñigo Goiri and Saeed Maleki and Ricardo Bianchini},
      year={2024},
      eprint={2311.18677},
      archivePrefix={arXiv},
      primaryClass={cs.AR},
      url={https://arxiv.org/abs/2311.18677}, 
}

@misc{mooncake,
      title={Mooncake: A KVCache-centric Disaggregated Architecture for LLM Serving}, 
      author={Ruoyu Qin and Zheming Li and Weiran He and Mingxing Zhang and Yongwei Wu and Weimin Zheng and Xinran Xu},
      year={2025},
      eprint={2407.00079},
      archivePrefix={arXiv},
      primaryClass={cs.DC},
      url={https://arxiv.org/abs/2407.00079}, 
}

@inproceedings{memory_metric_basic,
author = {Patel, Neel and Mamandipoor, Amin and Quinn, Derrick and Alian, Mohammad},
title = {XFM: Accelerated Software-Defined Far Memory},
year = {2023},
isbn = {9798400703294},
publisher = {Association for Computing Machinery},
address = {New York, NY, USA},
url = {https://doi.org/10.1145/3613424.3623776},
doi = {10.1145/3613424.3623776},
booktitle = {Proceedings of the 56th Annual IEEE/ACM International Symposium on Microarchitecture},
pages = {769–783},
numpages = {15},
location = {Toronto, ON, Canada},
series = {MICRO '23}
}

@misc{llmserving2,
      title={LLMServingSim 2.0: A Unified Simulator for Heterogeneous and Disaggregated LLM Serving Infrastructure}, 
      author={Jaehong Cho and Hyunmin Choi and Guseul Heo and Jongse Park},
      year={2026},
      eprint={2602.23036},
      archivePrefix={arXiv},
      primaryClass={cs.DC},
      url={https://arxiv.org/abs/2602.23036}, 
}

\end{document}